\documentclass[a4paper,11pt]{article}
\pdfoutput=1 

\usepackage{jcappub} 

\usepackage[T1]{fontenc} 
\usepackage[dvipsnames]{xcolor}
\usepackage{lscape}
\usepackage{adjustbox}
\usepackage[normalem]{ulem}
\def\be{\begin{equation}}
\def\ee{\end{equation}}
\def\CF3{{\sc cosmicflows-3}}

\usepackage{rotating}
\usepackage{multirow}
\usepackage{CJKutf8}
\usepackage{graphicx}	
\usepackage{amsmath}	
\usepackage{amssymb}	
\usepackage{newtxtext,newtxmath}

\usepackage[T1]{fontenc}
\usepackage{ae,aecompl}

\usepackage{color}
\usepackage{soul}

\usepackage{orcidlink}

\title{\boldmath Reconstructing the cosmological density and velocity fields from redshifted galaxy distributions using V-net}

\author[a,1~\orcidlink{0000-0001-7950-7864}]{Fei Qin,\note{Corresponding author.}}
\author[a,b]{David Parkinson,}
\author[a,b]{Sungwook E. Hong,}
\author[c]{Cristiano G. Sabiu.}

\affiliation[a]{Korea Astronomy and Space Science Institute,\\
776 Daedeok-daero, Yuseong-gu, Daejeon 34055, Republic of Korea}
\affiliation[b]{Astronomy Campus, University of Science and Technology,\\
776 Daedeok-daero, Yuseong-gu, Daejeon 34055, Republic of Korea}
\affiliation[c]{Natural Science Research Institute, University of Seoul,\\
163 Seoulsiripdaero, Dongdaemun-gu, Seoul 02504, Republic of Korea}

\emailAdd{feiqin@kasi.re.kr}
\emailAdd{davidparkinson@kasi.re.kr}
\emailAdd{swhong@kasi.re.kr}
\emailAdd{csabiu@uos.ac.kr}

\abstract{

The distribution of matter that is measured through galaxy redshift and peculiar velocity surveys can be harnessed to learn about the physics of dark matter, dark energy, and the nature of gravity. To improve our understanding of the matter of the Universe,  we can reconstruct the full density and velocity fields from the  galaxies that act as tracer particles. In this paper, we  use  
the simulated halos as proxies for the galaxies. We use a convolutional neural network, a V-net, trained on numerical simulations of structure formation to reconstruct the  density and velocity fields. We find that, with detailed tuning of the loss function, the V-net could produce better fits to the density field in the high-density and low-density regions, and improved predictions for  the probability distribution of  the amplitudes of the velocities.   However, the weights will reduce the precision of the estimated $\beta$ parameter. We also find that the redshift-space distortions of the  halo catalogue do not significantly contaminate the reconstructed real-space density and velocity field.  We  estimate the velocity field $\beta$ parameter by comparing the peculiar velocities of  halo catalogues to the reconstructed  velocity fields, and find the estimated $\beta$ values  agree with the fiducial value at the 68\% confidence level.    
}

\begin{document}
\maketitle
\flushbottom

\section{Introduction}
\label{sec:intro}

Observations of the Universe suggest that it is mostly composed of a mysterious dark matter, in addition to galaxies,  and that the expansion is currently being accelerated by the mysterious dark energy. Understanding the nature of dark matter and dark energy is crucial for us to comprehend the evolution of the Universe. This is also one of the key science goals of next-generation surveys, such as the Dark Energy Spectroscopic Instrument (DESI) \citep{DESI2016}, Euclid \citep{Percival2019}, the Widefield ASKAP L-band Legacy All-sky Blind Survey (WALLABY) \citep{Koribalski2020}, the Square Kilometre Array (SKA) \citep{2020PASA...37....7S}, and the Legacy Survey of Space and Time (LSST) \citep{Ivezi2008}.

However, observations of the Universe only provide us with information about galaxies, as non-luminous baryonic matter and dark matter are not directly detectable by telescopes. A commonly used method for leveraging this information to improve our understanding of the total matter in the Universe  is the reconstruction of the matter density and velocity fields using galaxy surveys, in which galaxies trace the underlying matter distribution of the Universe.

Peculiar velocities, which 
arise from the gravitational effects of matter density fluctuations, are good indicators of the matter density field in the nearby Universe.  
In linear perturbation theory, the relation between the peculiar velocity field of matter and  galaxy contrast field is given by \citep{Strauss1995,Springob2014,Springob2016,Carrick2015,
Boruah2020}
\be \label{betadef}
{\bf V}({\bf r}) = \frac{
\beta}{4\pi}\int \frac{{\bf r'}-{\bf r}}{|{\bf r'}-{\bf r}|} \delta_{\rm g}({\bf r'}) {\rm d}{\bf r'},
\ee 
with the velocity-density coupling parameter $\beta$ predicted to be
\begin{equation}
\beta = \frac{f}{b}\,.
\end{equation}
Here, 
$b\equiv\delta_{\rm g}/\delta$ is the biasing parameter, defined as the ratio between the galaxy contrast field $\delta_{\rm g}$ and  matter density contrast field $\delta$. $f(\Omega_{\rm m})$ is the growth rate of the large scale structure,  $f={\rm d} \ln \delta/ {\rm d}\ln a$, which is related to the matter density in standard Einstein gravity through the standard relation
\begin{equation}
f = \Omega_{\rm m} (z)^{0.545}\,.
\end{equation}

The parameter $\beta$ can be estimated using the so-called velocity-velocity comparison method, i.e., comparing the measured peculiar velocities to the (peculiar) velocity field 
reconstructed from the real-space density field, while this density field is  reconstructed from the redshift-space galaxy catalogues \citep{Strauss1995,Ma2012,Springob2014,Erdogdu2006a, Springob2014,Carrick2015,Springob2016,
Boruah2020,Lilow2021}. The measurement of $\beta$ parameter enables us to constrain the cosmological models. Comparing to the galaxy correlation function and power spectrum measurements,  
the velocity-velocity comparison is a much more  accurate method to derive $\beta$ since it is almost free from 
cosmic variance \citep{Carrick2015,Qin2019b,Said2020,Lilow2021}. Particularly, if the peculiar velocities measured from the Type Ia supernovae samples are used to compare to the reconstructions \citep{Carrick2015,Boruah2020}, this method will give an extraordinary accurate estimation of $\beta$, especially when the supernovae are too sparse to calculate the correlation functions and power spectrum very precisely.

In previous works \citep{Nusser1991,Zaroubi1995,Croft1997,Branchini1999,Kudlicki2000,Branchini2002,Erdogdu2006,Bilicki2008,Kitaura2012,Wang2012,Carrick2015}, the reconstruction of the velocity field from the density field is based on their relationship described by linear theory. However, this method does not predict very well the velocity field around very dense regions, mainly due to  complex nonlinear effects \citep{Wu2021}.  To explore the non-linear effects in reconstructions, several  methods have been developed, for example, the nonlinear forward modeling methods \cite{Seljak2017,Modi2018,Schmidt2019,Modi2021,Bayer2022}. Of closest similarity, \cite{Bayer2022} already performed the reconstruction of both density and velocity fields from a halo density field, which shares a similar goal to this paper. However, in addition to the method being applied, there are two main differences between \cite{Bayer2022} and our work. First, \cite{Bayer2022}   cuts at larger scales than considered in this work. Secondly, \cite{Bayer2022} reconstructs only the line-of-sight component of the velocity field, while we aim to reconstruct its full three-dimensional components.

Recently, artificial intelligence (AI) technology has been introduced into cosmology \citep[\& references therein]{Ravanbakhsh2017,Lucie-Smith2018,Merten2019,Carleo2019,Ntampaka2019}. 
Particularly, the utilization of convolutional neural networks  (CNN) for density- and velocity-field reconstruction \citep{Modi2018,He2019,Zhang2019,Mao2021,Pan2020,Wu2021,Hong2021,Ganeshaiah2022} has become a promising toolkit in the study of the large-scale structure (LSS) of the Universe.

In this paper,  we  use 
the (simulated) halos as proxies for the galaxies that will be observed (with no selection based on observability or mass). We will use simulations to train and test a novel CNN architecture, the V-net \citep{Milletari2016}, for reconstructing the real-space density fields from the halo catalogues of redshift-space, and reconstructing the velocity fields from the density fields.  In real surveys, the density field can help us to explore the inhomogeneous Malmquist bias. Therefore, in this paper, we intend to reconstruct both the density and velocity fields rather than only reconstructing the velocity field directly from halo fields. Although we do not need to study the Malmquist bias in this paper, we want to make our procedure closer to the cases of real surveys. We will also estimate the $\beta$ parameter by comparing the reconstructed velocity fields to the   halo  peculiar velocity catalogues, in order to explore whether a reasonable cosmological constraint can be obtained from the AI-predicted density and velocity fields.  We reconstruct the full three-dimensional velocity field because, although the upcoming galaxy surveys are sensitive only to radial velocities, it may be possible in the future to measure transverse velocities with the moving lens effect \citep{Hotinli2019,Hotinli2021}.

The paper is structured as follows. In Section~\ref{sec:datasim}, we introduce the simulation and the training data set, as well as the testing data set. In Section~\ref{sec:Unet}, we introduce the architecture of V-net. In Section~\ref{sec:recons}, we explore the density and velocity fields reconstruction. In Section~\ref{sec:beta}, we present the measurement of the $\beta$ parameter using this method and simulated data. A conclusion is presented in Section~\ref{sec:conc}.

\section{Creating the mock data}\label{sec:datasim}

\subsection{Particle simulations}

We generated our mock catalogues and density and velocity fields using \textsc{pinocchio} (the PINpointing Orbit Crossing Collapsed Hierarchical Objects code, \citep{2002MNRAS.331..587M,2002ApJ...564....8M,2002MNRAS.333..623T,2013MNRAS.433.2389M,2015MNRAS.452..686C,2017MNRAS.465.4658M,2017JCAP...01..008R,2016Galax...4...53M}). \textsc{pinocchio} is a semi-analytical algorithm that simulates structure formation through Lagrangian perturbation theory and ellipsoidal collapse, rather than a full $N$-body particle evolution approach used in codes such as \textsc{Gadget} \citep{SpringelVolker2001,SpringelVolker2005}. This approach does not track the full evolution of particles on an individual basis, but rather traces perturbations on different scales through a grid approach and identifies haloes from ellipsoidal collapse. This approach provides 
a massive speed-up in CPU time, allowing us to generate a large number of training sets very quickly.

To run the simulations, we chose values of the cosmological parameters given in Tab.~\ref{tab:cosmoparams}, which are taken from the maximum likelihood values estimated by the 9-year Wilkinson Microwave Anisotropy Probe (WMAP9) from \cite{2013ApJS..208...19H}. We assume a geometrically flat ($\Omega_k=0$) $\Lambda$CDM cosmology with the dark energy equation of state fixed to the cosmological constant value $w=-1$ and standard cold dark matter.

\begin{table}
\centering
\begin{tabular}{ l c c } \hline
 Parameter & Symbol & Value \\ \hline
 Matter density & $\Omega_{\rm m}$ & 0.25 \\  
 Cosmological constant density & $\Omega_\Lambda$ & 0.75 \\
 Baryon density & $\Omega_{\rm b}$ & 0.044\\
 Hubble parameter & $H_0$ & 70 km/s/Mpc \\
 Amplitude of fluctuations on scales of 8$h^{-1}$Mpc & $\sigma_8$ & 0.8 \\
 Spectral index of perturbation & $n_{\rm s}$ & 0.96 \\
 \hline
\end{tabular}
\caption{\label{tab:cosmoparams}The values of the cosmological parameters used for the \textsc{pinocchio} simulations.}
\end{table}

\begin{figure*} 
 \centering
\includegraphics[width=80mm]{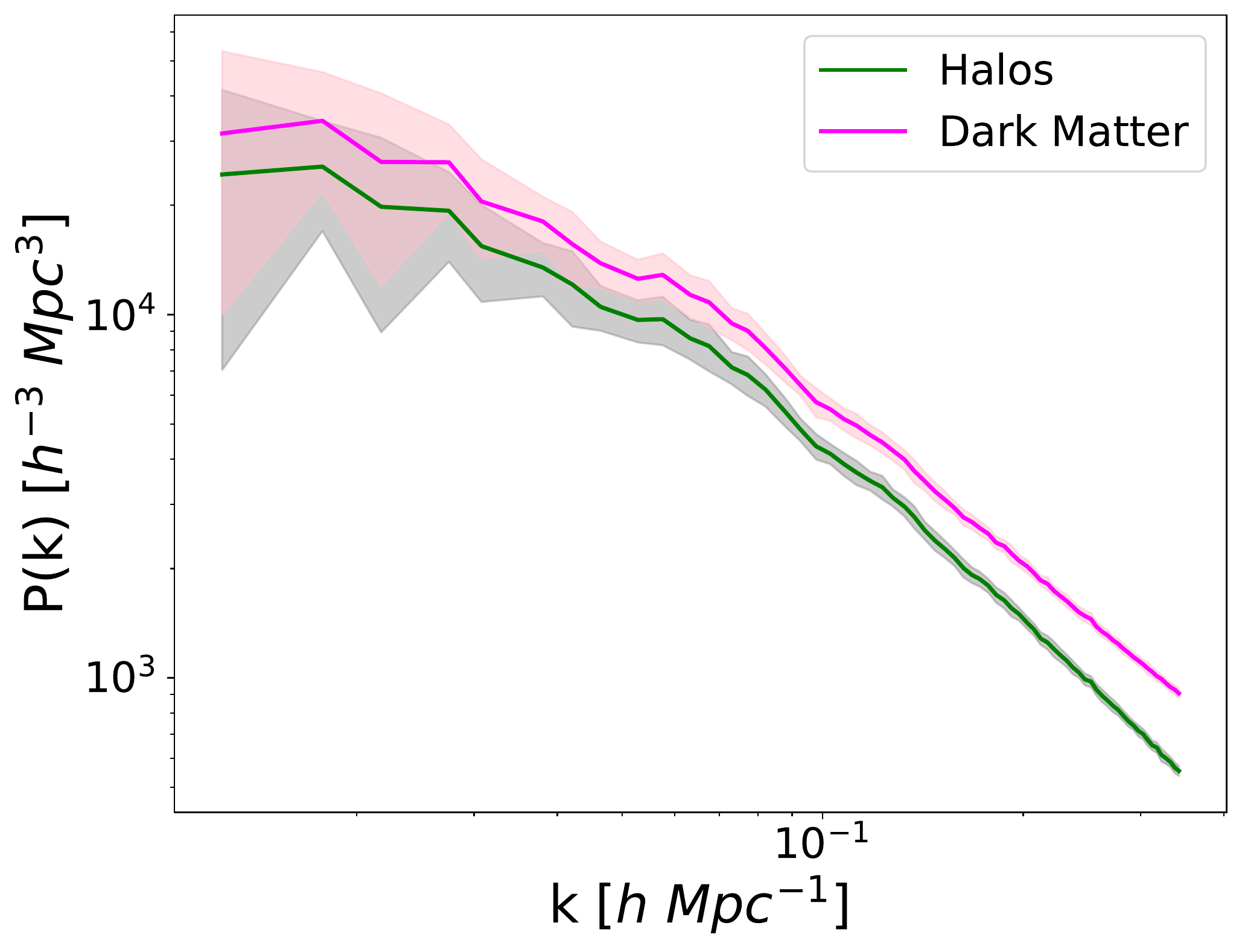} 
\caption{The density power spectrum of the simulations, measured in real space. 
The magenta curve is the average of the density power spectra of the dark matter of 42 simulation boxes. The pink area indicates the standard deviation  of the 42 measurements. 
The  green curve is the average of
the density power spectra of the  halos of the 42 simulation boxes. The grey area indicates the the standard deviation  of the 42 measurements.  
    }
\label{fig00}
\end{figure*}

The initial simulations are run with a cubic periodic box of side $L=500 h^{-1} \mathrm{Mpc}$, each containing $512^3$ dark matter particles. 
This gives a particle mass of $6.4\times 10^{10} h^{-1} {\rm M}_\odot$. We set a minimum halo member particle number 
of 10, meaning the minimum halo mass 
is $6.4\times 10^{11} h^{-1} {\rm M}_\odot$. 
We generated a set of 42 simulations, each with a different initial condition defined by a change in the random seed. This change in initial condition was the only difference between the boxes, as we generated simulations for only a single set of cosmological parameters for this training and test set. The real space power spectra for all of these 42 simulations is shown in Figure~\ref{fig00}. 

The output of each simulation is the position and velocity  of the particles, and a halo catalogue with position and velocity of each halo. All simulation 
outputs  are at redshift $z=0$, and we do not consider higher redshift matter distributions or construct a light cone that would include the effect of redshift evolution of the density field.  The halo catalogue is an automatic output of the \textsc{pinocchio} code through the \textit{fragmentation} algorithm, which mimics the hierarchical process of accretion of matter and merging of haloes \cite{2002MNRAS.331..587M}. In this work, we  use 
the haloes as proxies for the galaxies that will be observed, but with no selection based on observability or mass. Each simulation contains roughly 540,000 haloes, giving an average number density of $4.3\times 10^{-3} h^3 \mathrm{Mpc}^{-3}$, which is fairly close to the number density of the Two Micron All-Sky Survey (2MASS) Redshift survey \cite{1983ApJS...52...89H} (estimated to be $5.4\times 10^{-3} h^3 \mathrm{Mpc}^{-3}$). In Fig.~\ref{fig00}, the amplitude of the galaxy (halo) power spectrum is below that of the dark matter because the mean halo mass is small ($\sim 4.7\times 10^{12}h^{-1} {\rm M}_\odot$), leading to a halo bias less than unity \citep{Tinker2005,Basilakos2008,Qin2022}. 

\begin{figure*} 
 \centering
\includegraphics[width=80mm]{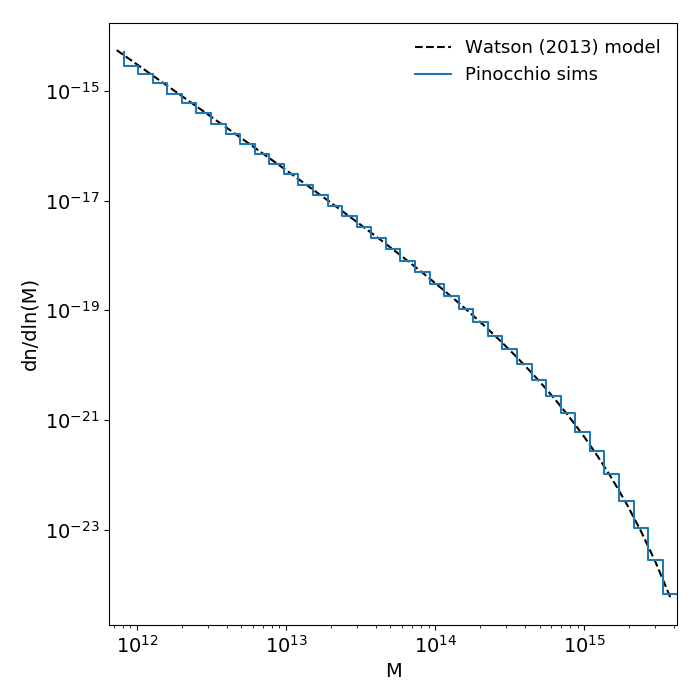} 
\caption{ The halo mass function. The mean simulation mass function is shown in the blue curve. The dash-doted curve is the  mass function model. }
\label{fig:mass_function_mean}
\end{figure*}

 To validate our minimum halo mass, we computed the mean of the mass function over the 42 simulations, and compared to a mass function model from \cite{Watson2013} using the \textsc{colussus} package \cite{2018ApJS..239...35D}. The mean simulation mass function is shown in Figure \ref{fig:mass_function_mean}. We find excellent agreement down to the minimum mass bin. 

\subsection{Training and testing data sets}\label{traintest}

To generate a large number of training cubes, each of the 42 large simulation boxes has been divided into $4^3$ sub-cubes. The density and velocity field of each sub-cube is pixelised with $32^3$ pixels. This pixel size is $\sim4~h^{-1}{\rm Mpc}$, chosen as the best compromise to predict velocities from the smooth density fields which are compared to measured peculiar velocities \cite{Berlind2000,Carrick2015,
Boruah2020,Springob2016}. We finally have 2,688 sub-cubes, each with side length of $125~h^{-1}{\rm Mpc}$. Around 20\% of these, i.e., 538 sub-cubes are randomly chosen as the testing data set. 

 \begin{figure*} 
 \centering
\includegraphics[width=60mm]{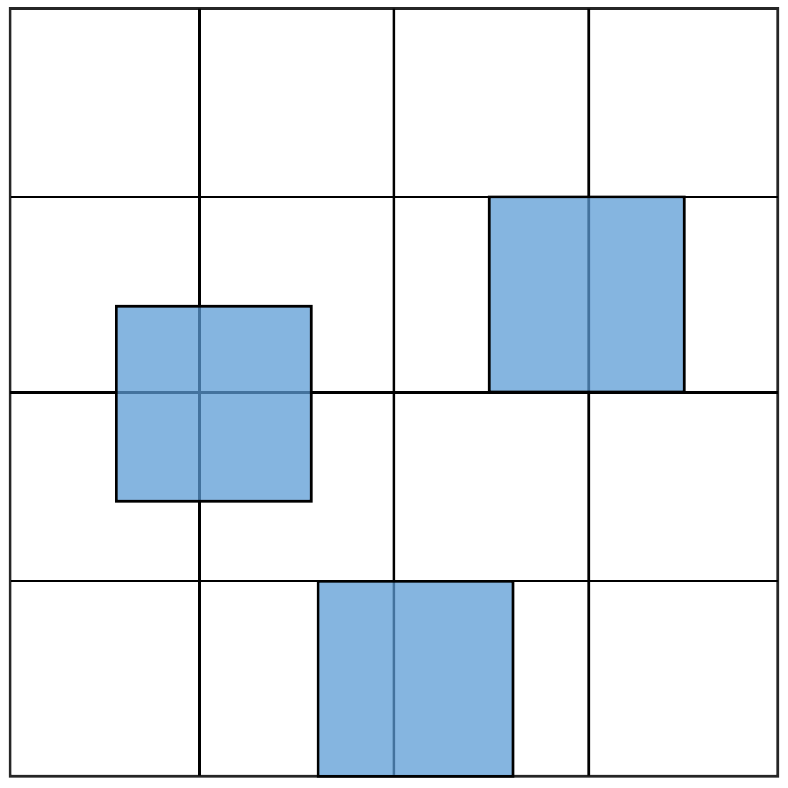}
\caption{The data augmentation used in this work. We randomly choose 6850 sub-cubes (blue squares)
which have overlap areas to the unattached 2,150 sub-cubes (white squares).}
\label{figdaug}
\end{figure*}

We applied data augmentation to the remaining 80\% of the sub-cubes to obtain 9,000 cubes as the training and validation data set.   As illustrated in Fig.~\ref{figdaug}, we randomly choose 6,850 sub-cubes which have overlap areas to the unattached 2,150 sub-cubes. The minimum stride to both the unattached sub-cubes and  other augmented sub-cubes is 8 pixels to avoid complete overlaps. Among these 9,000 sub-cubes, 20\% are used to validate CNN.  The number of each type and at each stage of pre-processing is given in Table \ref{tab:numsubcubes}.

\begin{table}
\centering
\begin{tabular}{l r } 
 \hline
 Dataset & Number  \\
 \hline
 All sub-cubes & 2,688 \\ 
 Testing set (no augmentation) & 538 \\
 Sub-cubes (for training \& validation) before augmentation & 2,150 \\
 Sub-cubes (for training \& validation) after augmentation & 9,000 \\
 Training set (augmented) & 7,200 \\
 Validation set (augmented) & 1,800 \\ 
 \hline
\end{tabular}
\caption{Number of sub-cubes of size 125 $h^{-1}$ Mpc used in each stage of CNN training, validation and testing.  }
\label{tab:numsubcubes}
\end{table}

 The density (contrast) field is defined as:
\be 
\delta({\bf r})=\frac{\rho({\bf r}) - \bar{\rho}}{ \bar\rho}
\ee
where $\rho({\bf r})$ is the mass density at real-space position ${\bf r}$,  $\bar\rho$ is the average mass density of the Universe.
The cloud-in-cell (CIC) algorithm\footnote{We use the \textsc{python} package \textsc{Nbodykit}  \url{https://github.com/bccp/nbodykit-cookbook/blob/master/recipes/painting.ipynb
}.} 
is adopted to the dark matter particles to obtain the density field cubes. 
The density values vary 
over three orders of magnitude, and so we renormalize the density field using 
\be  
\hat{\delta}= \ln(1+\rho/ \bar\rho)
\ee 
to reduce the variance.  

\begin{figure*} 
 \centering
\includegraphics[width=73mm]{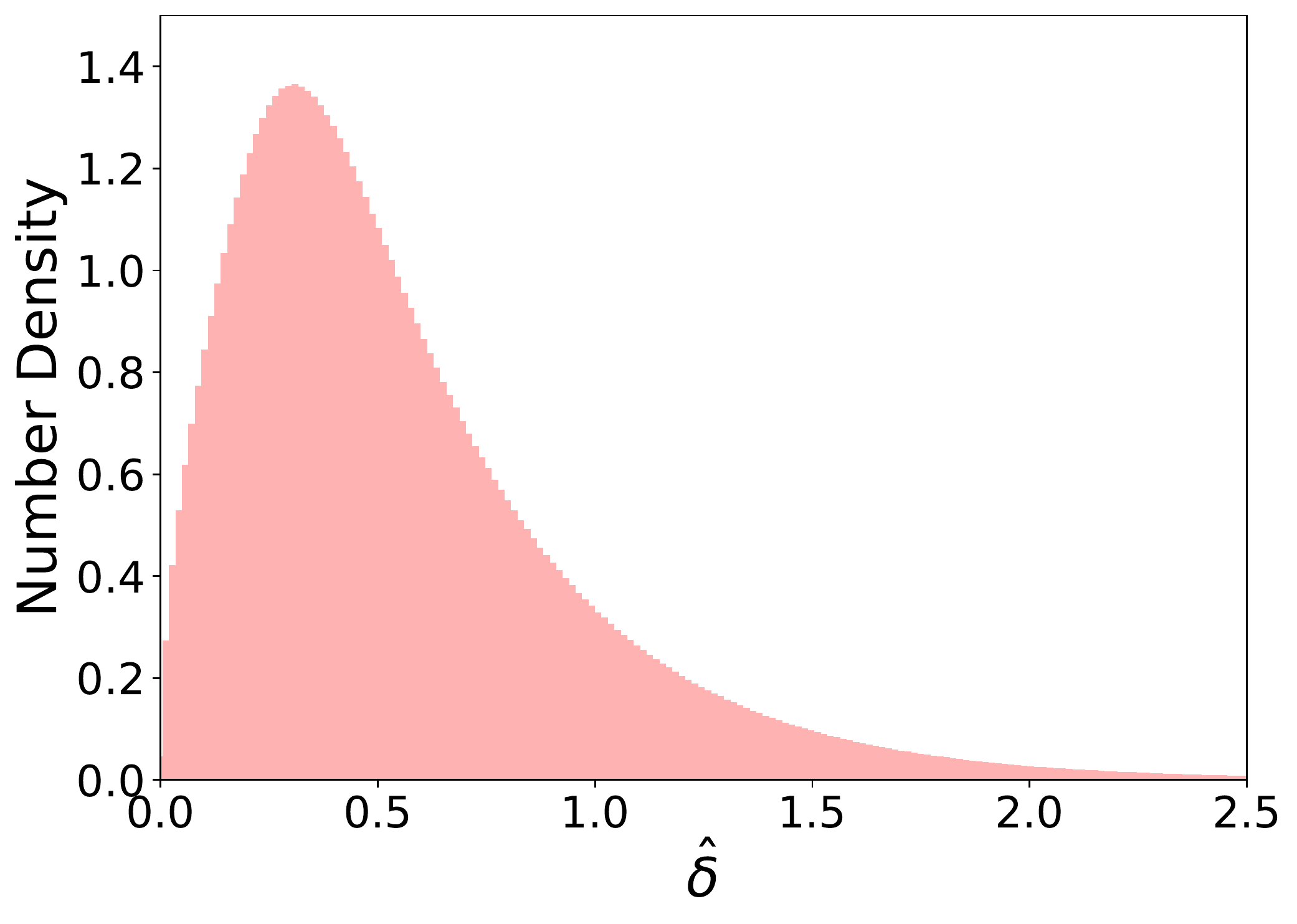}
\includegraphics[width=77mm]{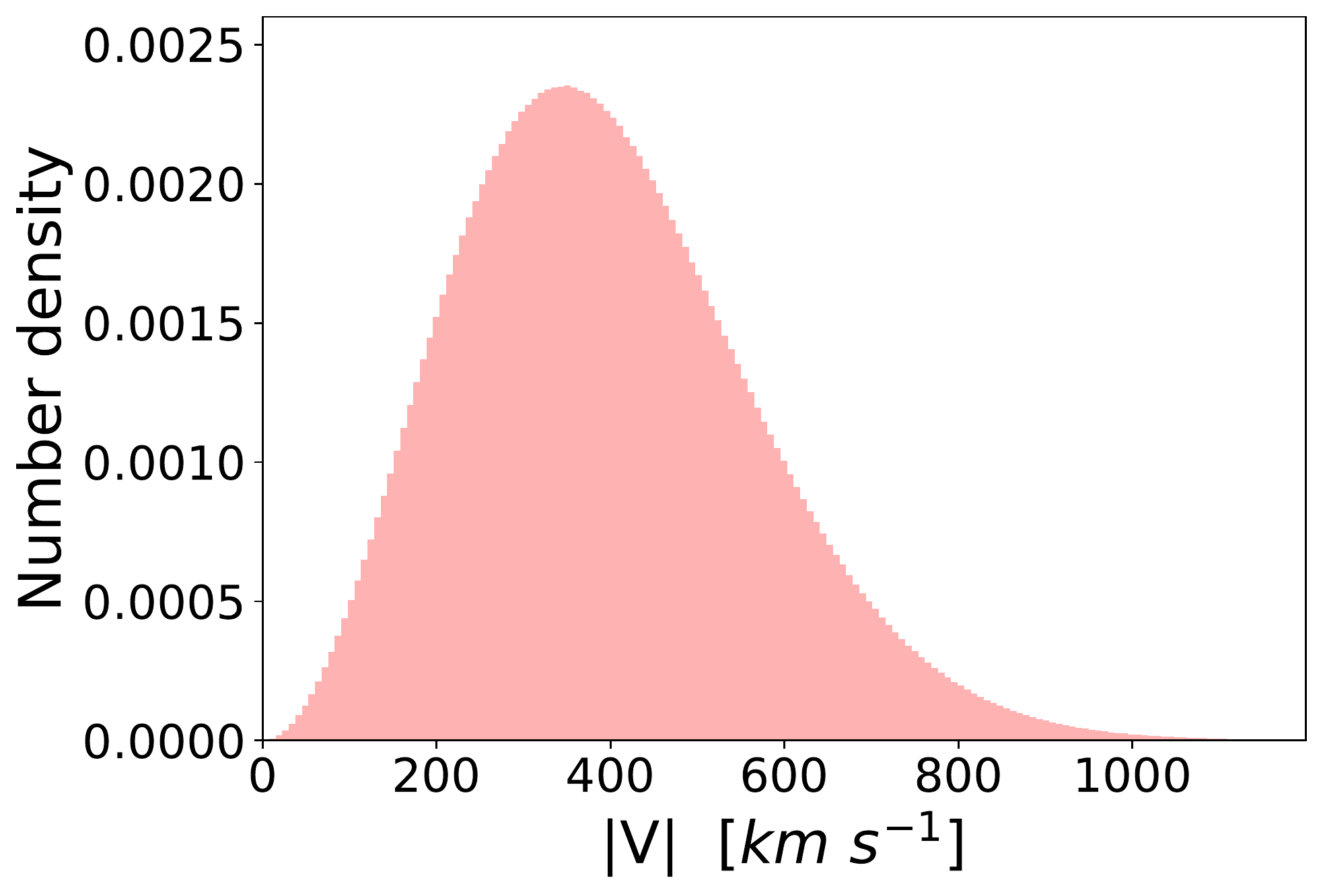} 
\caption{Probability distributions of  normalized density $\hat{\delta}$ (left) and the velocity amplitude $|\bf{V}|$ 
(right) in our simulation.}
\label{fig1}
\end{figure*}

For the velocity field cubes,  the velocities of each pixel
is simply calculated by averaging over the velocities of the particles in each pixel. We renormalize the velocity field using 
\be \label{normvel}
{\bf u }({\bf r}) = \frac{{\bf V({\bf r})}}{\sigma_v}
\ee 
so that the velocities will have the similar order of
magnitude as the density values.  The renormalization parameter $\sigma_v=250$ km s$^{-1}$  is the velocity dispersion (the standard deviation of the velocity amplitudes
) of our simulations. 
Fig.~\ref{fig1} shows the probability distributions of $\hat{\delta}$ and  $|\bf{V}|$  
in our simulations.

The halo contrast field cubes  are  obtained using  the same method as above, but this time using the halo catalogue instead of the dark matter particles. However, before this is done, 20\% of the halos are separated out to create the peculiar velocity catalogue, and these will be used only for estimating the $\beta$ parameter in the final stage. To obtain the redshift-space halo contrast field, redshift-space distortions (RSDs) have been added in the Z-direction  by adopting the distant-observer approximation, using\be 
Z_{\rm rsd} = Z_{\rm real}+\frac{1+z_{\rm h}}{H(z_{\rm h})}V_{\rm Z} ~,
\ee 
where $Z_{\rm rsd}$ and $Z_{\rm real}$ are the Z-direction coordinates of halos in redshift and real space, respectively.\footnote{Note that such a distant-observer approximation is valid only when the comoving distance at the given redshift is significantly larger than the box size. For our box sixe with $125~h^{-1}{\rm Mpc}$, it corresponds to $z_{\rm h} \gtrsim 0.2$. However, in this paper we focus on the feasibility of our V-net reconstruction for the $\beta$-measurement, rather than the application to any observational data. We leave applying the proper RSDs along the line-of-sight at low-redshifts as future works.} $V_{\rm Z}$ is the Z-direction component of the halo velocity, $z_{\rm h}$ is the cosmological redshift of our simulation snapshot\footnote{If using a simulation of lightcone, $z_{\rm h}$ is the redshift of each halo.}, and $H(z_{\rm h})$ is the Hubble parameter.

\subsection{Halo peculiar velocity catalogues}

Once we obtain the reconstructed {\it real-space} density and velocity fields from the redshift-space halo contrast field, we can compare the reconstructions to the measured peculiar velocities to estimate the cosmological parameter $\beta$. In observations, the measured peculiar velocity values are mainly obtained from galaxies using the Tully-Fisher relation \citep{Tully1977,Hong2019} or Fundamental Plane \citep{Djorgovski1987,Howlett2022}. In this paper we generate halo peculiar velocity catalogues in order to estimate $\beta$ through the velocity-velocity comparison method.

To build the  halo peculiar velocity  catalogues, we assign cosmological redshifts $z_{\rm h}$  to the halos based on their radial real-space comoving distances $r$  using 
\be\label{Dz}
r=\frac{c}{H_0}\int_0^{z_{\rm h}}\frac{dz'}{\sqrt{\Omega_{\rm m} (1+z')^3+\Omega_{\Lambda}}},
\ee
although these halos are sub-sampled from the halo catalogues of the snapshots corresponding to cosmological redshift zero. The line-of-sight peculiar velocity of a halo is calculated using 
\be 
s_{\rm g} ={\bf V}_{\rm g} \cdot \hat{{\bf r}} ~,
\ee 
where ${\bf V}_{\rm g}$ is the velocity of halos 
known from the simulations. $\hat{{\bf r}}$ is the unit vector corresponding to the real-space position ${\bf r}$, rather than setting the Z-direction as the line-of-sight. Then, the observed redshift of a halo is given by \cite{Colless2001,Hui2006,Davis2014,Scrimgeour2016,Qin2021a,Qin2021b}
\be \label{travp}
1+z_{\rm obs} = (1+z_{\rm h})\left(1+\frac{s_{\rm g}}{c}\right)~,
\ee
where $c$ is the speed of light. The $z_{\rm obs}$ values will be used to estimate $\beta$ in Section~\ref{sec:beta}.

\section{Machine learning architecture: V-net}\label{sec:Unet}

The machine learning U-net architecture was initially proposed by \cite{Unet2015}. Overall, this method has a similar architecture to the convolutional autoencoder (CAE).
Both U-net and CAE consist of two stages --- first, the encoding stage extracts features from the original image by increasing the number of channels and decreasing the image size of each channel.
Then follows the decoding stage, which reconstructs the new image by decreasing the number of channels and increasing the image size of each channel.
In addition to CAE, U-net  concatenates both the outputs from the encoding stage and inputs from the decoding stage for the convolution to generate the outputs at the decoding stage, which prevents the loss of small-scale features in the CAE.

\begin{figure*}[bt] 
\centering
 \includegraphics[width=\columnwidth]{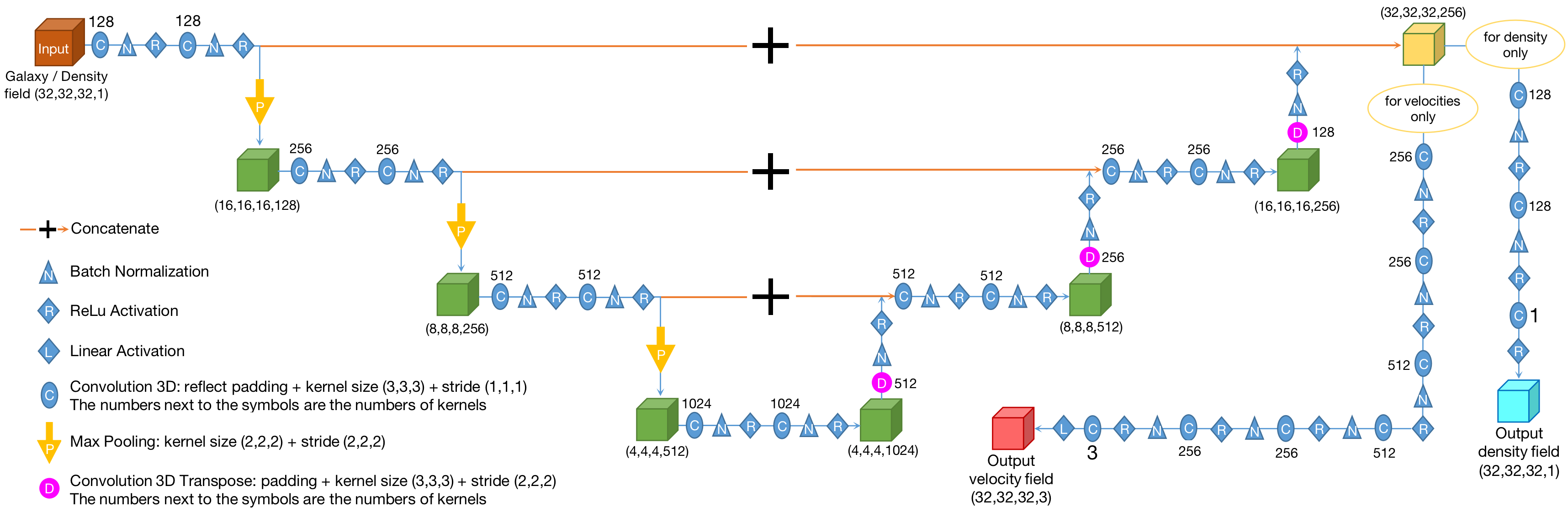}
 \caption{The V-net   
 architecture used to reconstruct the density and velocity fields in this work .} 
 \label{fig1unet}
\end{figure*}

Recently, a three-dimensional extension of U-net, referred to as V-net, was proposed by \cite{Milletari2016}. The implementation of V-net has been utilized in the reconstruction of LSS in various studies, including \cite{Zhang2019,Hong2021,Wu2021}. The present study employs a CNN architecture, as depicted in Fig.~\ref{fig1unet}, which is based on the original V-net framework with modifications inspired by the works of \cite{Hong2021,Wu2021}.

To undertake the density field reconstruction, the redshift-space halo contrast fields are set to be the inputs of the CNN architecture. The real-space density fields are set to be the outputs of the architecture. Each input halo contrast field has spatial dimension $32^3$ and one channel, i.e., it has a shape of $(32,~32,~32,~1)$. 
\begin{enumerate}
\item First, the input halo contrast fields are fed into two convolution layers. Each convolution layer is designed to have 128 convolution kernels with a shape of $3^3$ and a scanning stride of 1. 
Reflect padding, which adds the padding pixels to the boundaries by mirror-reflecting the inner pixels, is applied to ensure that the outputs have the same dimension as the inputs, as well as to minimize the loss of information around the boundaries. A batch normalization \cite{Ioffe2015} and a rectified linear unit (ReLU, \cite{Glorot2011}) activation function are used following each convolution layer. The final outputs of this step are a set of feature fields, and    each feature field has a shape of $(32,~32,~32,~128)$.

\item Then, these feature fields are fed into a max-pooling layer with a shape of  $2^3$ and 2-stride. Only the maximum values from nearby $2^3$ pixels pass through the max-pooling layer, and the dimension of each feature field has been reduced to $(16,~16,~16,~128)$ in this step.  
\item For further feature extraction and compression,  the outputs of the above step are processed by the same operations as steps~1--2 three more times. However, the number of convolution kernels used in each operation are changed to 256, 512, and 1024, and there is no max-pooling used in the third operation,  as shown in the Fig.~\ref{fig1unet}. The final output of this step has a shape of $(4,~4,~4,~1024)$. 
\item Then, they are fed into a de-convolution (or transpose convolution) layer. This layer is designed with reflect padding and 512 de-convolution kernels.   Each de-convolution kernel has a shape of $3^3$ and the scanning stride is 2. 
A batch normalization and a ReLu activation function are applied. The final output has a shape of $(8,~8,~8,~512)$.  
\item The outputs of the above `decoding' step will be concatenated with the feature fields of the equivalent (with a size of (8,~8,~8,~512)) `encoding' stage. Then they will be fed into two convolution layers which are the same as Step~1, but using 512 convolution kernels. 
\item The outputs of the above step are processed using the same operations as Steps~4--5 twice, but using 256 and 128 kernels, respectively. There is no convolution layer used in the last operation. The final output of this step has a shape of $(32,32,32,256)$.   
\item To obtain the reconstructed density field, the feature fields are further processed by two convolution layers of 128 kernels. The kernel has a shape of $3^3$ and 1-stride. 
The padding, batch normalization and ReLU activation function are applied to the outputs of each convolution layer too. The final convolution layer has only one kernel and followed by a ReLU activation function.
The shape of the convolution kernel is $3^3$ and the stride is 1.      
\item Finally, we obtain the density field reconstructed from the halo catalogues. 
To reconstruct the velocity field, the reconstructed density fields  (rather than the halo contrast fields) are set to be the inputs of the CNN architecture. Each density field has a shape of $(32,~32,~32,~1)$ too.  Correspondingly, the outputs of Step~6 will be processed by six convolution layers.
The numbers of kernels used in the layers are listed in Fig.~\ref{fig1unet}.  The final convolution layer has three kernels and followed by a linear activation function.
The shape of each convolution kernel is $3^3$ and the stride is 1.        
\end{enumerate}

We implement the \textsc{TensorFlow}\footnote{\textsc{TensorFlow}:  \url{https://www.tensorflow.org/}} package to build the above architecture. The optimization method is the Adam algorithm \cite{Kingma2014}.
The mini-batch size is 32. We stop training if the validation loss reaches its minimum and tend to be flat.

\section{Reconstructing the density and velocity fields}\label{sec:recons}

\subsection{Reconstructing the density field from the halo catalogue}
\label{sec:recons_dens}

We use the 9,000 training/validation halo cubes in redshift space  and the density field cubes in real space as the inputs and outputs
, respectively, in order to train the CNN. The trained V-net, corresponding to the minimum validation loss, is then applied to the 538 testing halo cubes to obtain the reconstructed density fields of the real space.  We define the loss function as
\be \label{lossden}
\mathcal{L}_{\hat{\delta}} 
=\frac{1}{N_{\rm pix}}\sum^{N_{\rm pix}}_{i=1} \left( w_{{\rm \hat{\delta}},i} \left|\hat{\delta}_{{\rm p},i}-\hat{\delta}_{{\rm t},i} \right| \right) ~,
\ee 
where $\hat{\delta}_{{\rm p},i}$, $\hat{\delta}_{{\rm t},i}$, and $w_{\hat{\delta},i}$ are the prediction and true values of the normalize density, and the weight factor for the $i$-th pixel, respectively.

\begin{figure} 
\centering
\includegraphics[width=76mm]{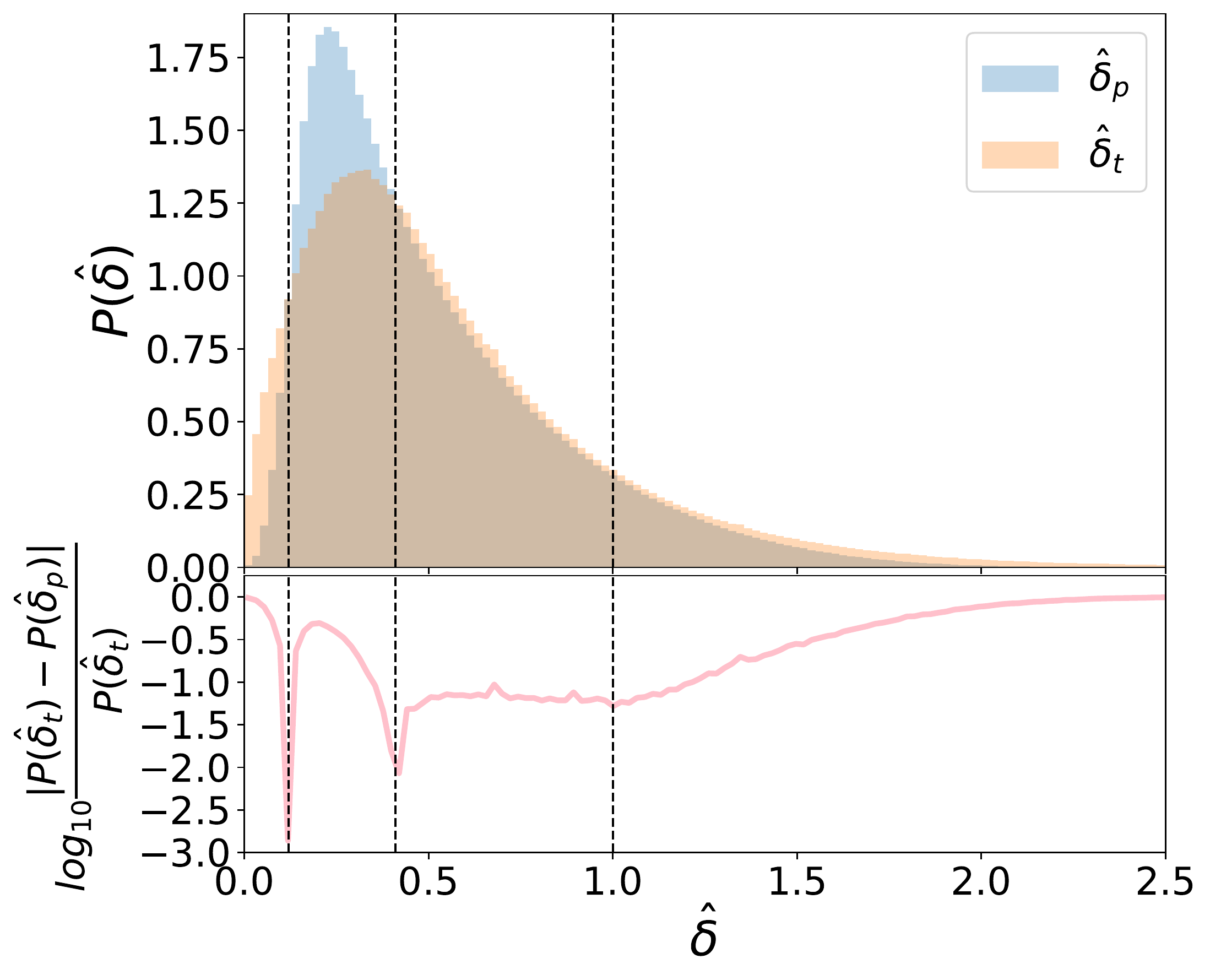}
\includegraphics[width=74mm]{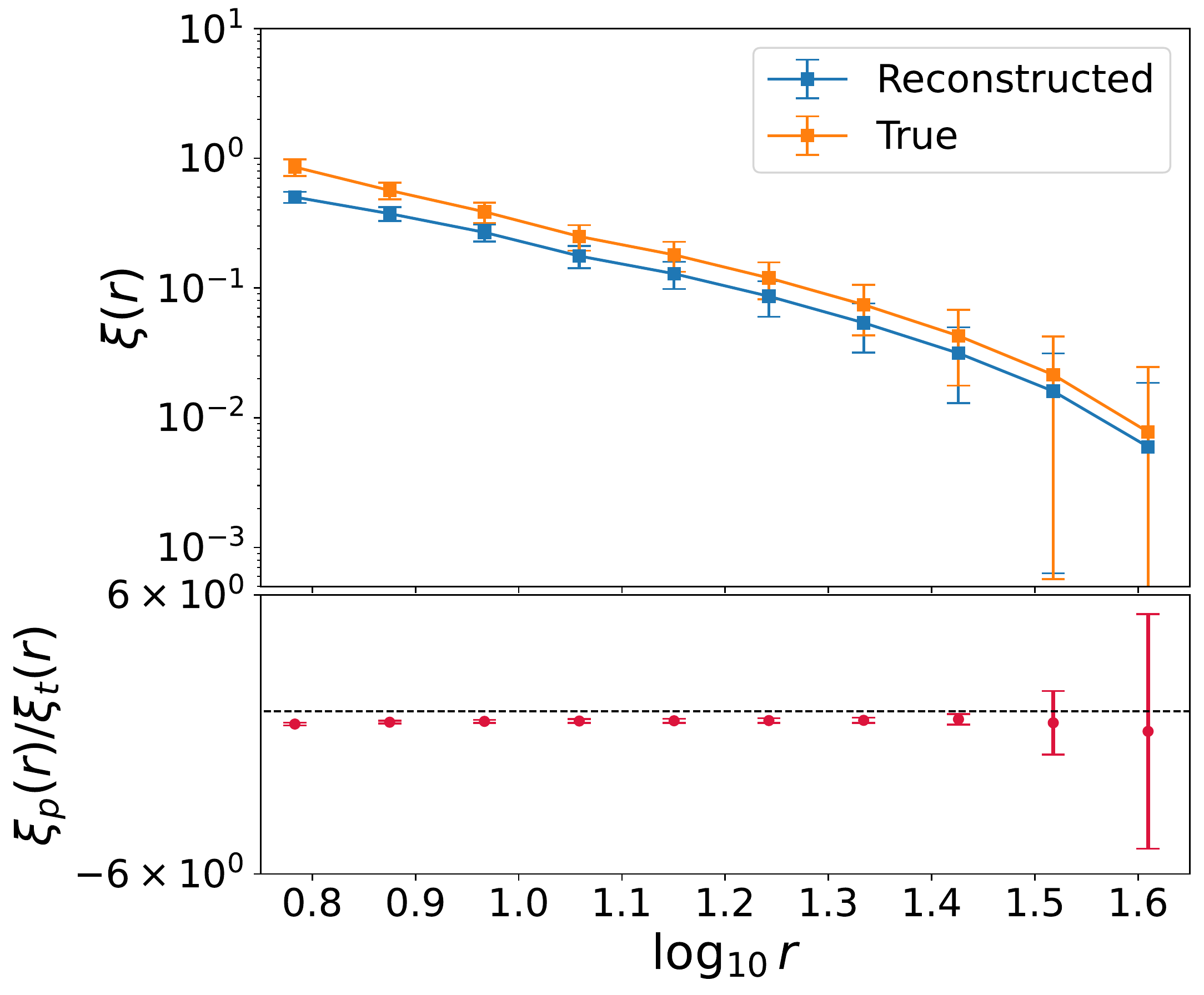}
 \caption{  Left:  Comparing the reconstructed density field to the true density field by setting the weights $w_{{\rm \hat{\delta}},i} =1$ in the loss function Eq.~\ref{lossden}. 
  The top panel shows
  the probability distributions of the reconstructed density values $\hat{\delta}_{\rm p}$ (blue bars) and true density values $\hat{\delta}_{\rm t}$ (orange bars) for 538 testing cubes, respectively.  
  The bottom panel shows the difference between the true density values and reconstructed density values. The dashed vertical lines indicate the positions of the minimums of the pink curve: 
$\hat{\delta}_{\rm t}=0.12$; $\hat{\delta}_{\rm t} =0.41$; $\hat{\delta}_{\rm t}=1$.  Right:  Two-point correlation functions of the reconstructed (blue) and true (orange) density fields for the averaging of 538 testing cubes, respectively. The unit of $r$ is $h^{-1} {\rm Mpc}$.
 }
 \label{fighistw1}  
\end{figure}


\begin{figure*} 
\centering
  \includegraphics[width=49mm]{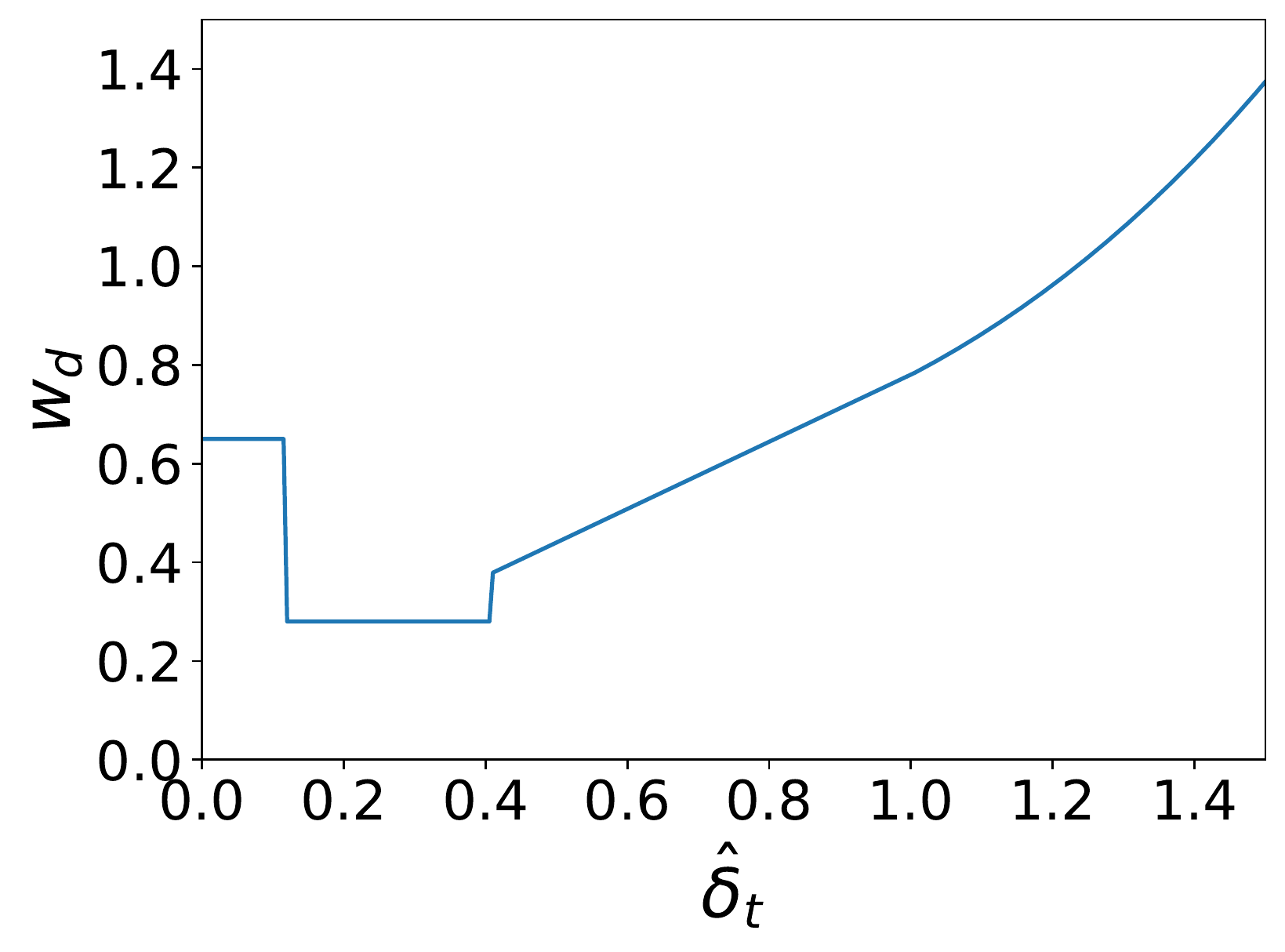}
  \includegraphics[width=48mm]{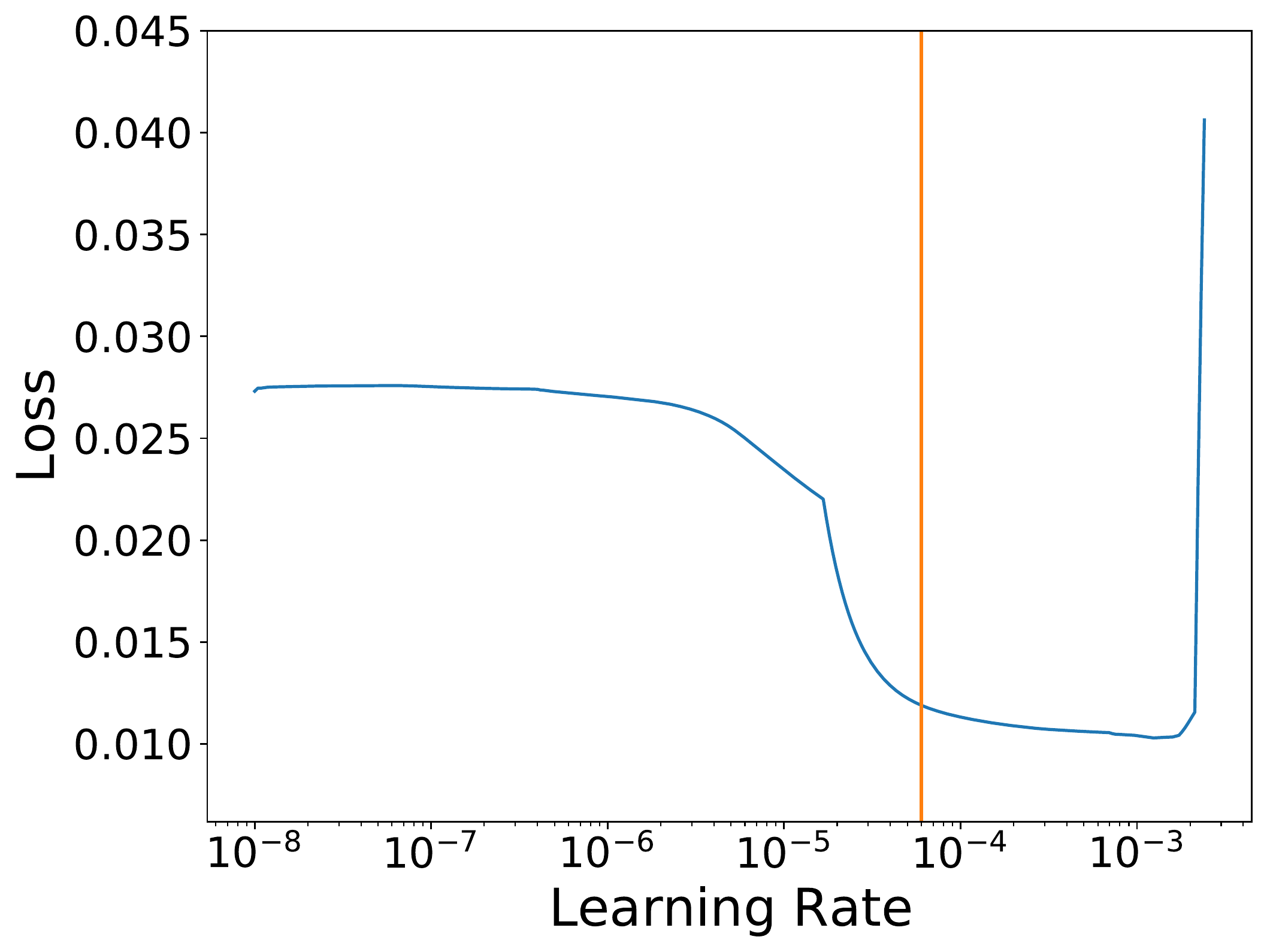} 
  \includegraphics[width=51mm]{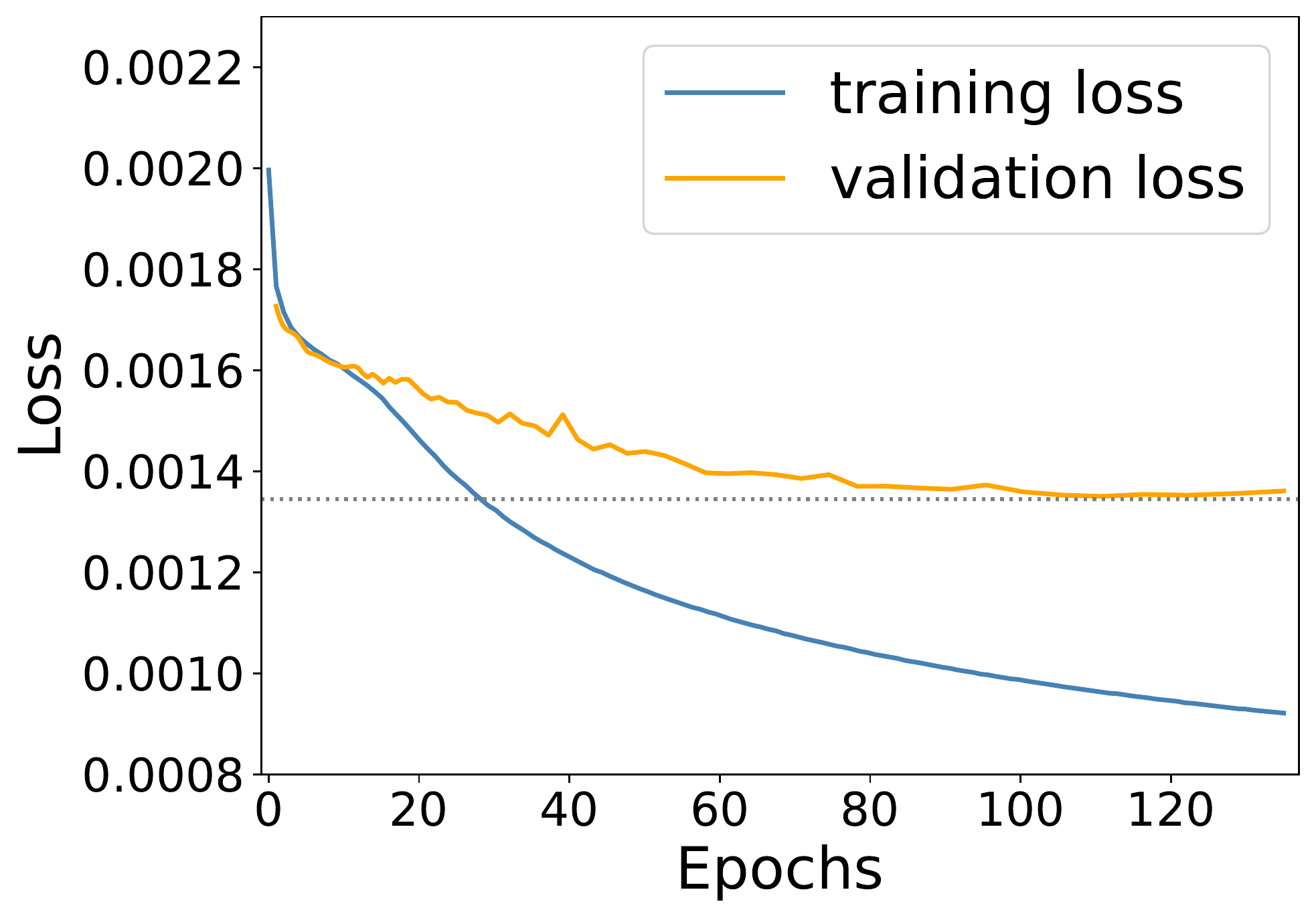}
 \caption{ 
 Left: Weights for reconstructing the density field, $w_{\hat{\delta}}$ as a function of true density field.
 Middle: Evolution of loss function as a function of learning rate (blue curve) and the learning rate we have adopted (orange line).
 Right: Loss function of density field reconstruction as a function of training epoch.
 }
 \label{figdensweightloss}
\end{figure*}

\begin{figure*} 
\centering
  \includegraphics[width=\columnwidth]{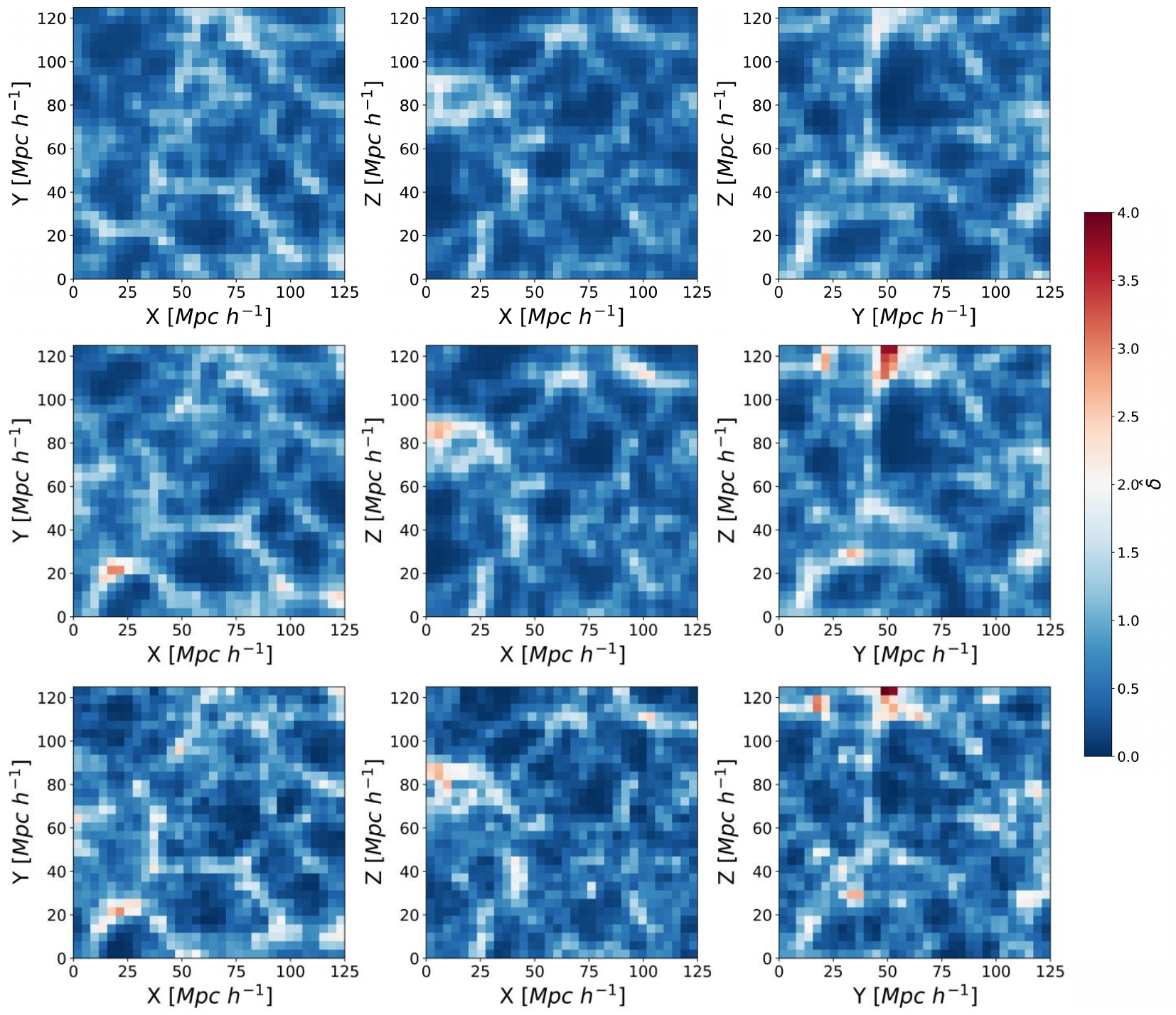}
 \caption{Comparison between the reconstructed density fields (top and middle) and  the true density fields (bottom) in XY-, XZ- and YZ-planes, respectively.  In top panels, the density fields are reconstructed by setting the weights $w_{{\rm \hat{\delta}},i} =1$  in the loss function. In the middle panels, the density fields are reconstructed by setting the weights using Eq.\ref{field} in the loss function.   
 The thickness of each slice is 4 $h^{-1}$ Mpc  (one pixel size).  
 }
 \label{fig5denrecon}
\end{figure*}

We initially test the case of reconstructed the density field with unity weights, setting $w_{{\rm \hat{\delta}},i} =1$ in Eq.\ref{lossden}. The top-left panel of Fig.\ref{fighistw1} shows the comparison between the reconstructed density field and true density field in this case.  In the bottom-left panel of Fig.~\ref{fighistw1}, the pink curve shows the difference between the true and reconstructed density values, and there are three minimum points in the pink curve at $\hat{\delta}_{\rm t}=0.12$; $\hat{\delta}_{\rm t} =0.41$; $\hat{\delta}_{\rm t}=1$. In the top-right panel of Fig.\ref{fighistw1}, the orange 
curve and the error bars are 
the average and the standard deviation of the two-point correlation function 
of the 538 true density fields, respectively, 
while the blue curve and the error bars are
the average and the standard deviation of the correlation function of the 538 reconstructed density fields, respectively.
The two-point correlation function measured from the reconstructions (blue curve) is  lower than the measurement from the true density fields (orange curve). This is due to the V-net failing to predict the higher density regions, therefore, the reconstructed density field do not have enough clusters, as shown in the top panels of Fig.\ref{fig5denrecon}. In addition, from Fig.\ref{fig5denrecon}, we find that the voids generated by reconstruction are simpler compared to the true density fields.

Overall, we find that the lack of a weight factor $w_{\hat{\delta},i}$ different to one results in an inability for the V-net to accurately reproduce the density field in both high- and low-density regions. This is likely caused by the imbalanced training data, where the model may overfit 
to the most frequently occurring class (pixel) with $\hat{\delta} \sim 0.3$ ($\rho \sim \bar{\rho}$), leading to poor performance on under-represented classes in both extremely dense and under-dense regions (see the left-top panel of Fig.~\ref{fighistw1}).  
Therefore,  the inclusion of weight factors is crucial for improving predictions in both high- and low-density regions by adjusting for the imbalance in the training data.


For matching the distribution of the reconstructed density values ${\delta}_{\rm p}$ to the true density values ${\delta}_{\rm t}$,  we need to weight the contributions to the loss from different density regions. We choose the four regions that bounded by  the minimum points of the pink curve of Fig.\ref{fighistw1}. The simplest way would be to use four different constants to weight  the four regions, respectively. However, as illustrated in  the pink curve of Fig.\ref{fighistw1}, the difference between the reconstructed and true distribution of densities $\log_{10}|(P(\hat{\delta}_{\rm t})-P(\hat{\delta}_{\rm p}))/P(\hat{\delta}_{\rm t})|$ 
is gradually  increasing, starting from $\hat{\delta}_{{\rm t},i}>0.41$, due to the V-net failing to predict the density values of denser region.   Therefore, we can choose the expression for $w_{\hat{\delta},i}$ to be density dependent for density values much larger than the mean. We use a linear function of $\hat{\delta}_{{\rm t}}$ in the range $0.41 >\hat{\delta}_{{\rm t},i}<1$, and a power-law function of $\hat{\delta}_{{\rm t}}$  for $\hat{\delta}_{{\rm t},i}>1$. 

We define the functional form of the weights as
\be\label{field}
w_{\hat{\delta},i}=\left\{
\begin{aligned}
&a_1, &0< &\hat{\delta}_{{\rm t},i}\leq 0.12\\
&a_2, &0.12< &\hat{\delta}_{{\rm t},i}\leq 0.41 \\
&a_3\hat{\delta}_{{\rm t},i}+a_4, &0.41<&\hat{\delta}_{{\rm t},i}\leq1 \\
&a_5\hat{\delta}_{{\rm t},i}^{a_6}+b, &&\hat{\delta}_{{\rm t},i}>1 \\
\end{aligned}
\right. ,
\ee
where the parameter $b$ is to keep the weights continuous at $\hat{\delta}_{{\rm t},i}=1$.  
In the above expression, there are six free
parameters $a_1$, $a_2$, $a_3$, $a_4$, $a_5$ and $a_6$. By simultaneously matching the probability distributions of the reconstructed density values to the true density values in the four regions, we find the optimal choice of the parameters are:   
\be\label{field0}
a_1=0.65,~~a_2=0.28,~~a_3=0.68,~~a_4=0.1,~~a_5=0.25,~~a_6=3.
\ee
and use the continuity constraint to calculate $b=0.53$. The final weight function of Eq.\ref{field} is shown in the left-side panel of Fig.~\ref{figdensweightloss}.


The blue curve in the middle panel of Fig.~\ref{figdensweightloss} shows the evolution of loss function as a function of learning rate of the Adam optimizer. If the learning rate is lower than $10^{-6}$, the update of the parameters is too slow, which is presented as a flat slope of the blue curve. On the contrary, 
a too high learning rate ($>2\times10^{-3}$) prevents finding a solution. Therefore, we choose the learning rate $6\times10^{-5}$ for density field reconstruction.   
The right panel of Fig.~\ref{figdensweightloss} shows the training loss (blue curve) and validation loss (orange curve) with the learning rate $6 \times 10^{-5}$. Beyond epoch equal to 90, the validation loss tends to be flat, therefore, we stop training at epoch equal to 135.

Fig.~\ref{fig5denrecon} shows the comparison between the reconstructed density field images (reconstructed with weights, middle panels) and  the true density field images (bottom panels) in the XY-, XZ- and YZ-planes. The V-net predictions of tiny and large filaments, as well as the voids, match the true fields with high fidelity by adding weights to loss function. Also, the V-net gives reasonable predictions of the density fields with less contamination from the RSDs (applied to the Z-direction) of the halo catalogues. Note that, however, the reconstructed filamentary structure is smoother and less complex than the true field, overall. Also, for some very massive clusters, our reconstruction exhibits somewhat different substructures than the true field (e.g., see YZ-plane with $(Y, Z) = (40~h^{-1}{\rm Mpc}$, $120~h^{-1}{\rm Mpc})$).
We suspect this may be partially due to the limitation of our halo sample, which has a similar number density to the 2MASS Redshift survey. Structures associated with much smaller scales than the scales allowed by our halo sample  may not be reconstructed well even with our CNN method. Therefore, to marginalize such problems, one may need to try either adopting galaxy/halo samples with higher number density or making the spatial resolution of the density cube lower. We leave both possibilities as future works.

\begin{figure*} 
\centering
  \includegraphics[width=76mm]{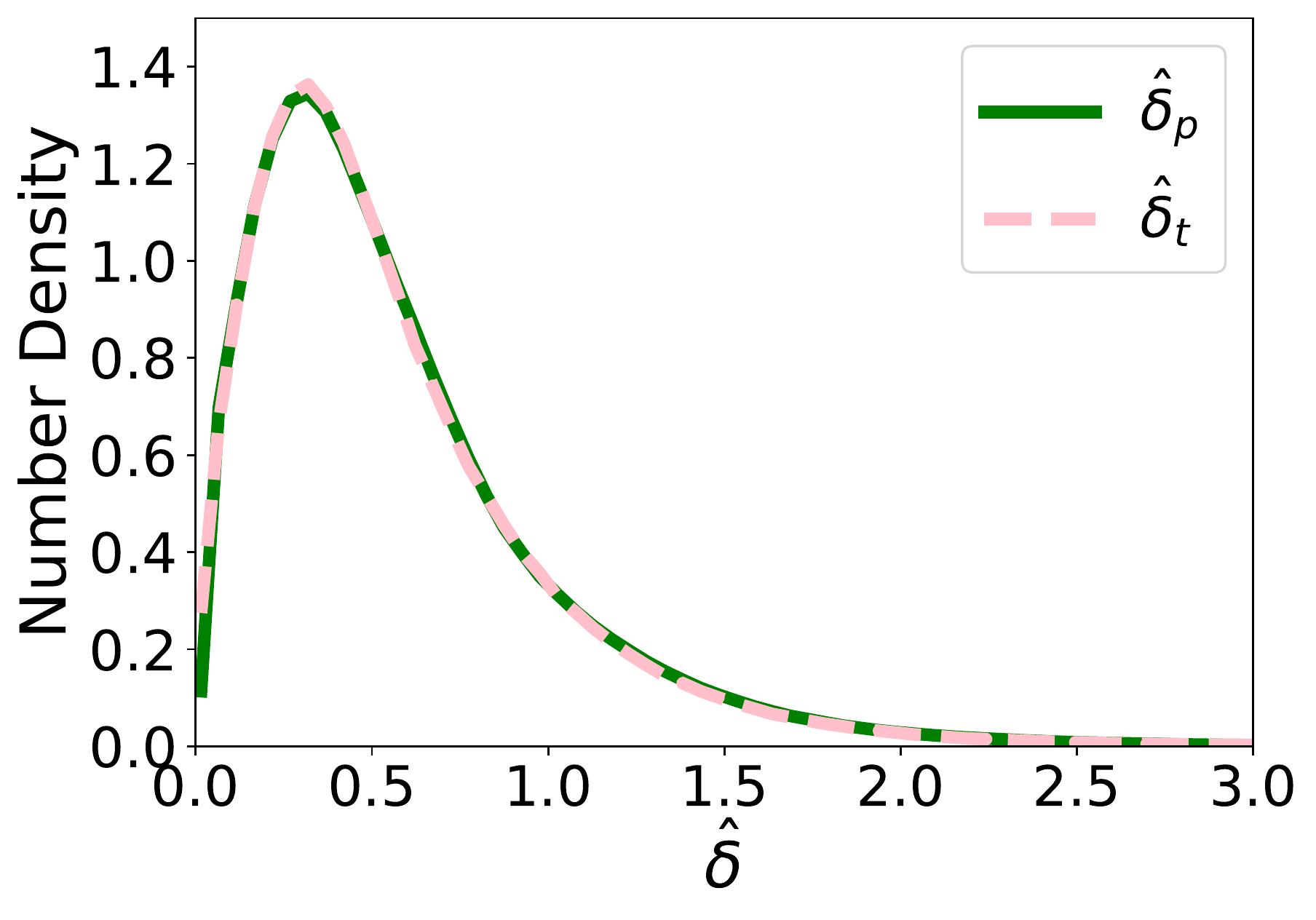}\includegraphics[width=74mm]{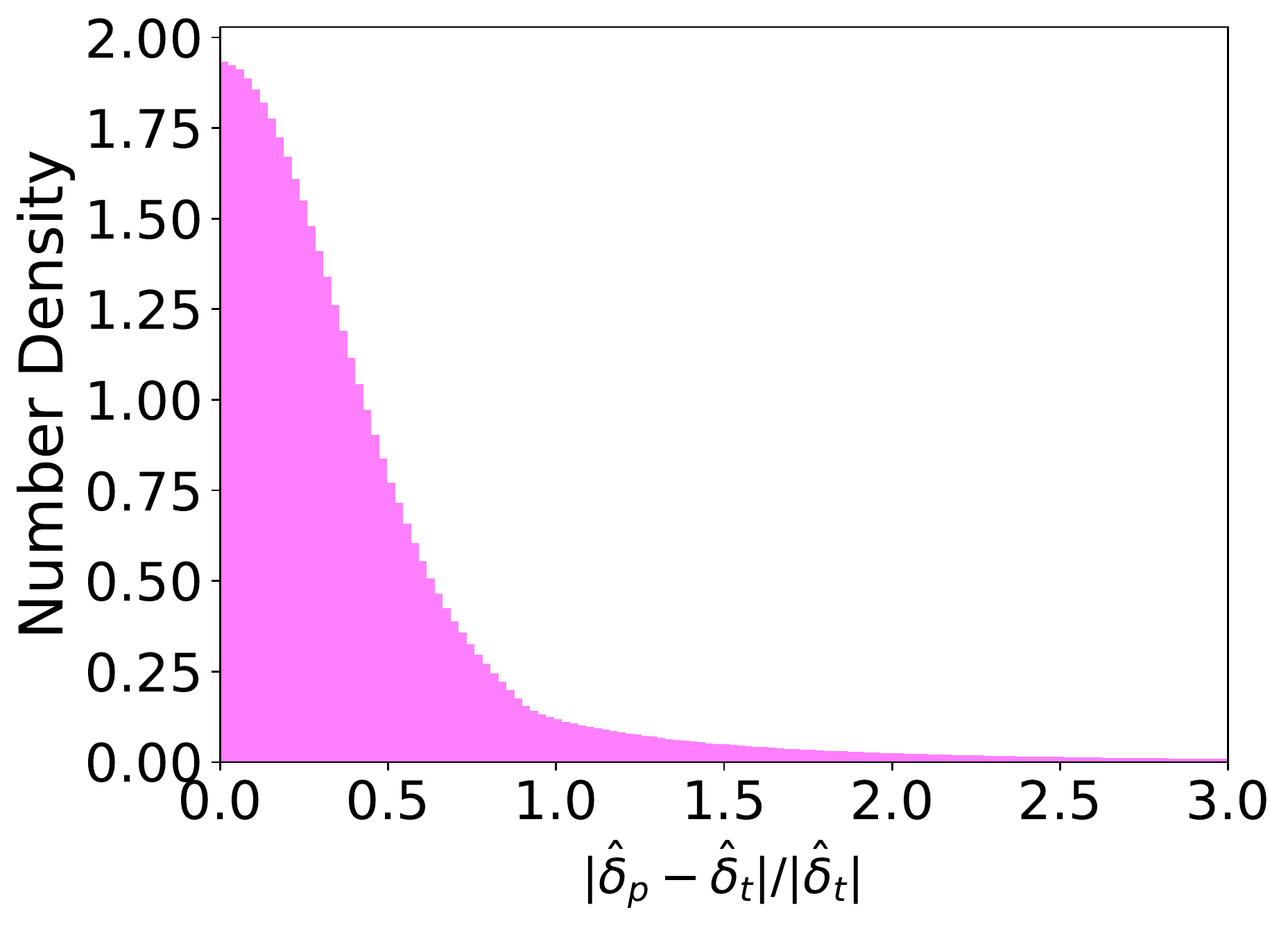}
  \includegraphics[width=79mm]{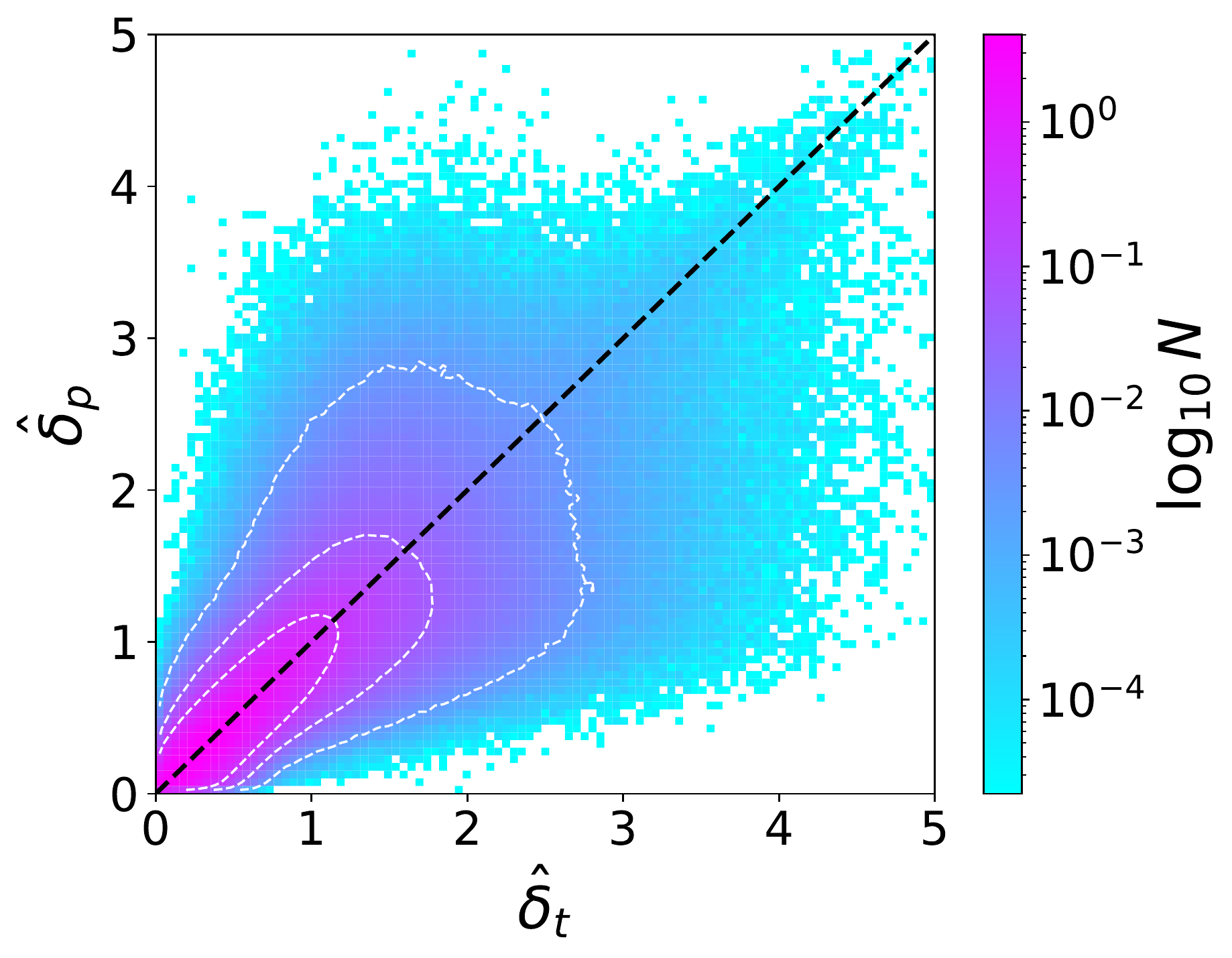}
  \includegraphics[width=71mm]{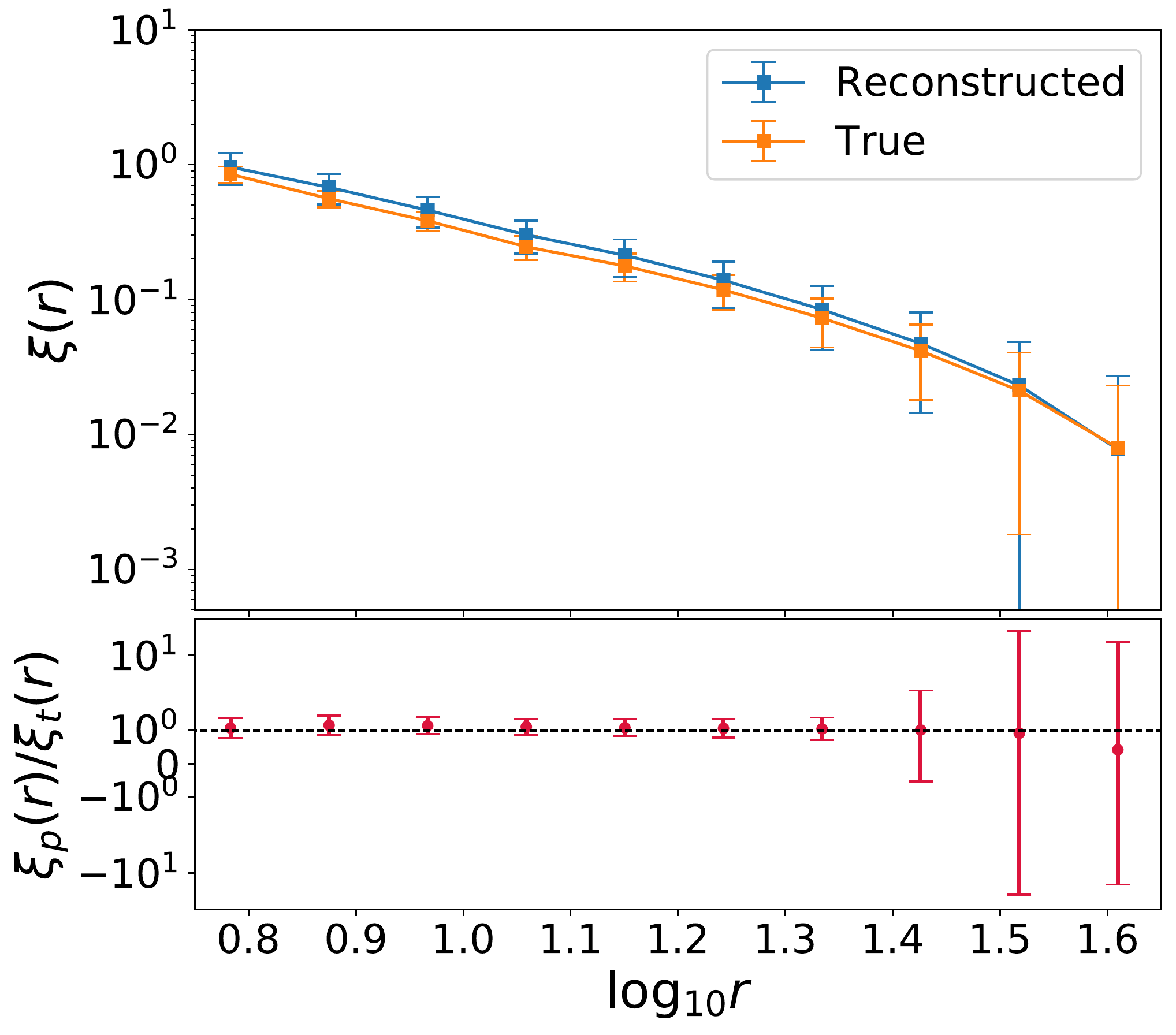}
 \caption{ 
 Top-left: Probability distributions of the reconstructed density values $\hat{\delta}_{\rm p}$ (green solid curve) and true density values $\hat{\delta}_{\rm t}$ (pink dashed curve) for 538 testing cubes, respectively. 
 Top-right: Probability distribution of residual between the reconstructed and true density values for these cubes.  
 Bottom-left: The reconstructed density values against the true values, with the white colored contours indicating
 the 68\%, 95\%, and 99.7\% confidence 
 levels.  
 Bottom-right: Two-point correlation functions of the reconstructed (blue) and true (orange) density fields for the averaging of 538 testing cubes, respectively. The unit of $r$ is $h^{-1} {\rm Mpc}$.}
 \label{figdenscomp}
\end{figure*}

In the top-left panel of Fig.~\ref{figdenscomp}, the distributions of the reconstructed density values $\hat{\delta}_{\rm p}$ of the 538 testing cubes is shown in green curve, while the corresponding true density values $\hat{\delta}_{\rm t}$ is shown in pink curve. 
The fractional residual between $\hat{\delta}_{\rm p}$ and $\hat{\delta}_{\rm t}$ is mostly 
smaller than 1, as displayed in the top-right panel. 
The bottom-left panel shows $\hat{\delta}_{\rm p}$ against $\hat{\delta}_{\rm t}$, with the white colored contours indicating the 68\%, 95\%, and 99.7\%
confidence levels. The reconstructed density values largely agree with the true density values. 
Note that there exist a small population with very high true density value and relatively low reconstructed value ($\hat{\delta}_{\rm p} \lesssim 3-4 \lesssim \hat{\delta}_{\rm t}; \rho_{\rm p} \lesssim 10^{3-4} \bar{\rho} \lesssim \rho_{\rm t}$), while there is no such population with very high reconstructed density value and relatively low true value ($\rho_{\rm t} \lesssim 10^{3-4} \bar{\rho} \lesssim \rho_{\rm p}$).
This matches well with our observation in Fig.~\ref{figdenscomp} that the reconstructed density field is somewhat smoother than the true density field.

In the bottom-right panel, the orange 
curve and the error bars are 
the average and the standard deviation of the two-point correlation function 
of the 538 true density fields, respectively, while the blue curve and the error bars are the average and the standard deviation of the correlation function of the 538 reconstructed density fields, respectively. As expected from the above analyses, both two-point correlation functions match well with each other within the error bar. The error bars of the orange curve represent the variance between the sub-cubes and the whole simulation box, i.e., cosmic variance. The error bars of the blue-colored curve represent the cosmic variance combined with the errors caused by the CNN reconstruction.

\begin{figure} 
\centering
\includegraphics[width=85mm]{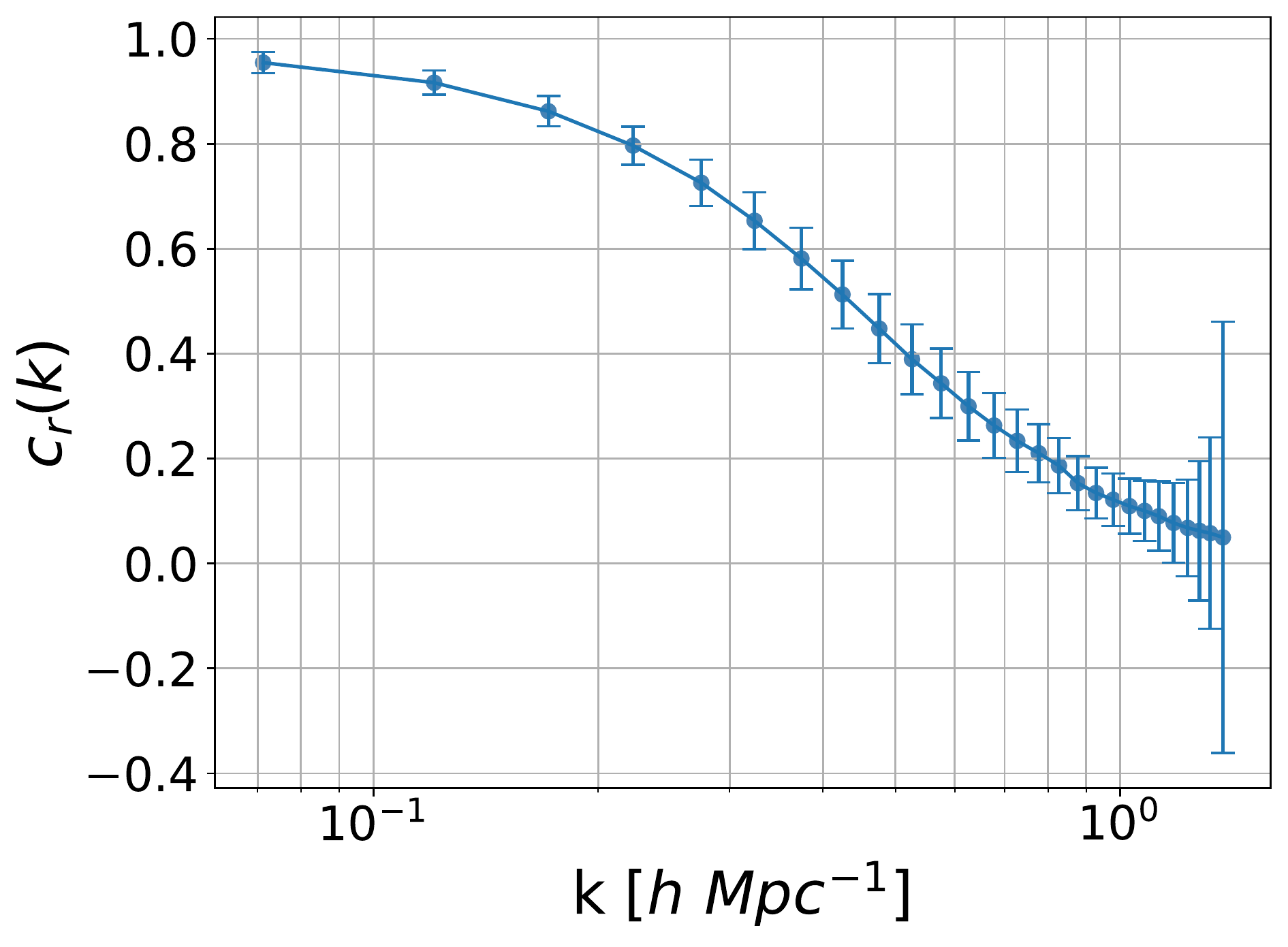}
 \caption{ The cross correlation coefficient between the power spectrum of the  538 true and 538 reconstructed density fields. The blue dots and error bars are the average and the standard deviation of the 538 measurements, respectively.}
 \label{figcr}  
\end{figure}

The cross-correlation coefficient between the power spectrum of the true and reconstructed density values is defined as:
\be
c_{\rm r}(k)=\frac{P_{\rm tp}(k)}{\sqrt{P_{\rm tt}(k)P_{\rm pp}(k)}}
\ee 
where $P_{\rm tp}(k)$ is the cross power spectrum of the true and reconstructed density fields, 
$P_{\rm tt}(k)$ (and $P_{\rm pp}(k)$) is the auto power spectrum of the true (and reconstructed) density fields. As shown in Fig.\ref{figcr}, the reconstructed density field is about
80\% correlated with the true density at $k$= 0.2 $h$ Mpc$^{-1}$ and about 50\%
correlated at $k$= 0.45 $h$ Mpc$^{-1}$.

\subsection{Reconstructing the velocity field from the reconstructed density field}
\label{sec:recons_vel}

To reconstruct the velocity fields from the density fields, we first need to apply the V-net model trained from the above density field reconstruction to the 9,000 training halo contrast cubes to obtain 9,000 reconstructed density cubes. After that, we use these reconstructed density fields as the inputs to train the V-net. The V-net architecture 
for velocity reconstruction, which is slightly different from density field reconstruction, is presented in Section~\ref{sec:Unet} and Fig.~\ref{fig1unet}. The corresponding 9,000 true velocity field cubes, normalized using Eq.~\ref{normvel}, are set to be the outputs of the V-net for training. Then, the trained V-net with the minimum validation loss is applied to the 538 reconstructed test density fields to obtain the reconstructed test velocity fields.   

Following the arguments in \cite{Wu2021} and \cite{Wu2023}, we define the loss function as:
\be \label{lossvel}
\mathcal{L}_{u,\theta} = \mathcal{L}_u + \mathcal{L_\theta}
=\frac{1}{N_{\rm pix}}\sum^{N_{\rm pix}}_{i=1} \left( w_{u,i} \left| u_{{\rm p},i}-u_{{\rm t},i} \right| +  \left|\cos\theta_{i}-1 \right| \right) ~ ,
\ee  
where $u=|{\bf u}|$ denotes the amplitude of the normalized velocity, $\cos\theta$ is the angle between the predicted velocity and true velocity, and   $w_{u}$ is the weight factors for $u$, respectively. Defining the loss function using the velocity components $u_x$, $u_y$ and $u_z$ will introduce too many degrees of freedom in the weight function, as a result, it will be too hard to determine the form of the weight function.  In addition,  we aim to reconstruct $\beta$ from our velocity field reconstruction. As written in Eq.\ref{betswei}, the $s$ parameter used in estimation
of $\beta$ parameter depends on ${\bf V}_{\rm p}\cdot\hat{\bf r}=|{\bf V}\cos\alpha|$, where $\alpha$ is the angle between velocity field and line-of-sight direction.
This equation indicates that  proper reconstruction of magnitude is crucial for  proper estimation of $\beta$ parameter. Correct reconstruction of the angle is important, but that is the angle relative to the observer. By splitting the loss function into magnitude and angle, we can apply a weight to the magnitude only, which we found to be most effective in precisely measuring $\beta$. 


\begin{figure} 
\centering
\includegraphics[width=85mm]{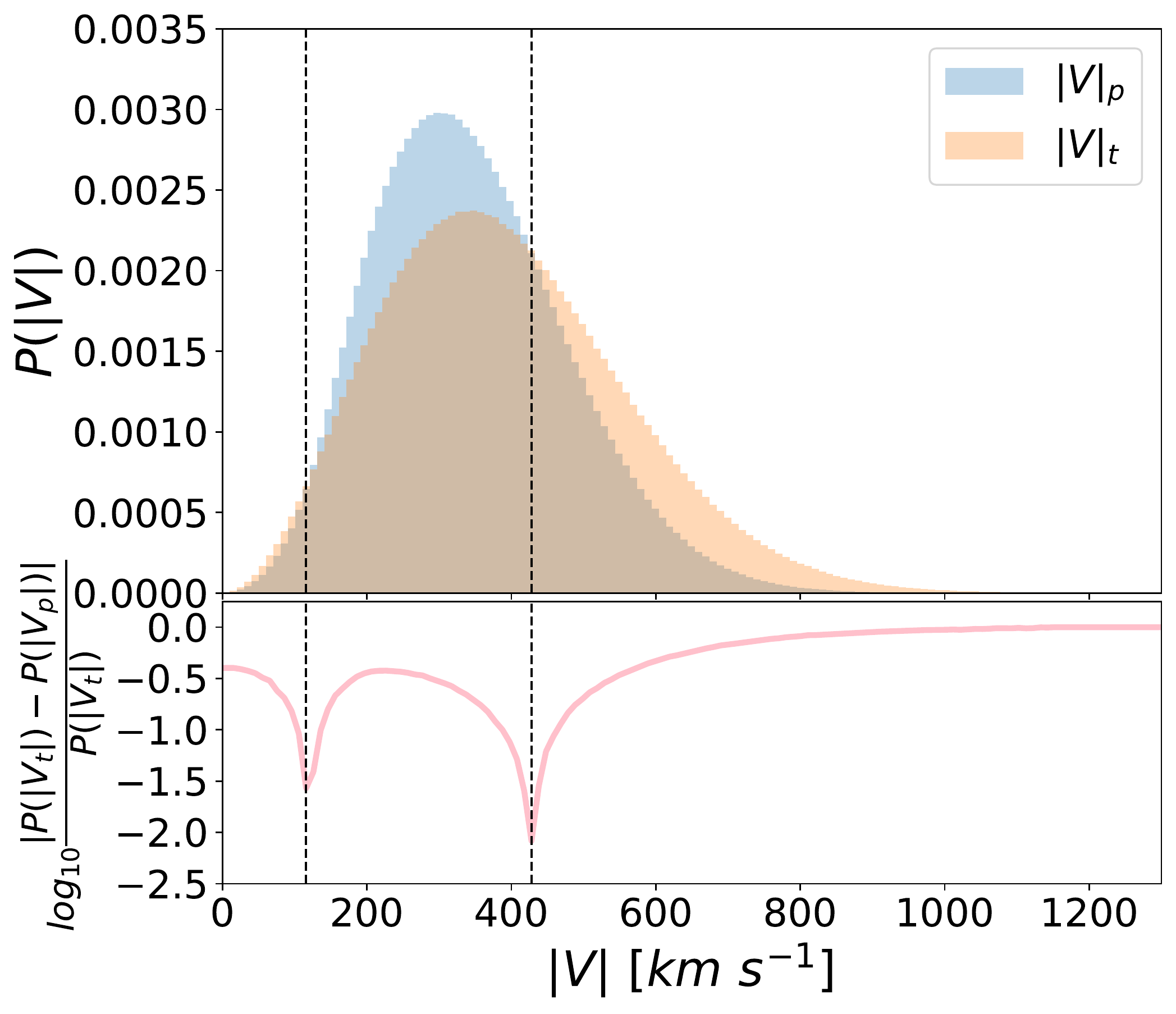}
 \caption{Comparing the reconstructed velocity field to the true velocity field by setting the weights $w_{{\rm \hat{u}},i} =1$. 
  The top panel shows
  the probability distributions of the amplitude of the  reconstructed velocities $|{\bf V}_{\rm p}|$ (blue bars) and the amplitude of the true velocities $|{\bf V}_{\rm t}|$ (orange bars) for 538 testing cubes, respectively.  
  The bottom panel shows the difference between the true values and reconstructed values. The dashed vertical lines indicate the positions of the minimums of the pink curve: 
$|{\bf V}_{\rm t}|=115.77$ km s$^{-1}$ and 
$427.85$ km s$^{-1}$.}
 \label{fighistw2}  
\end{figure}

\begin{figure*}
\centering
  \includegraphics[width=51mm]{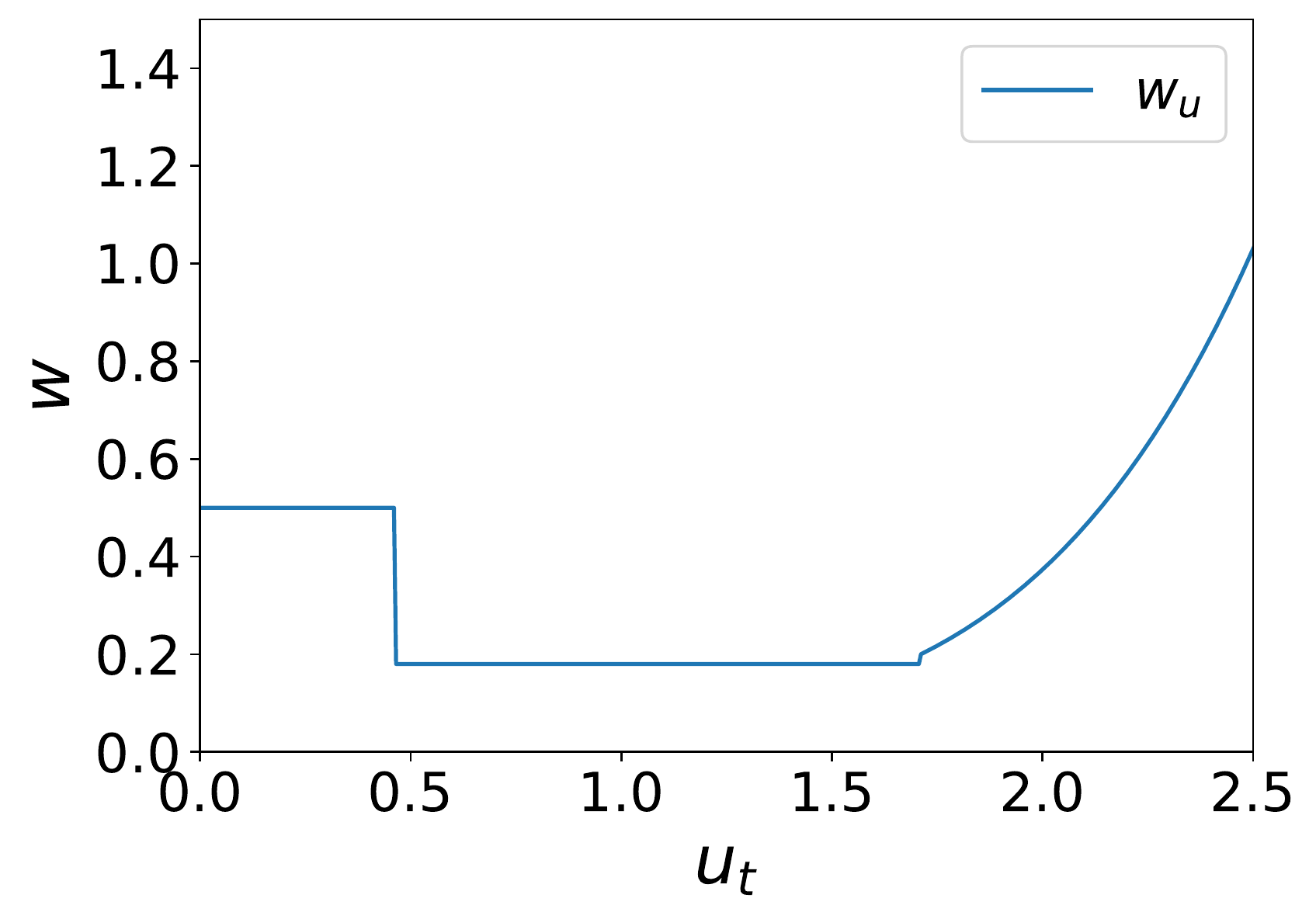}
  \includegraphics[width=50mm]{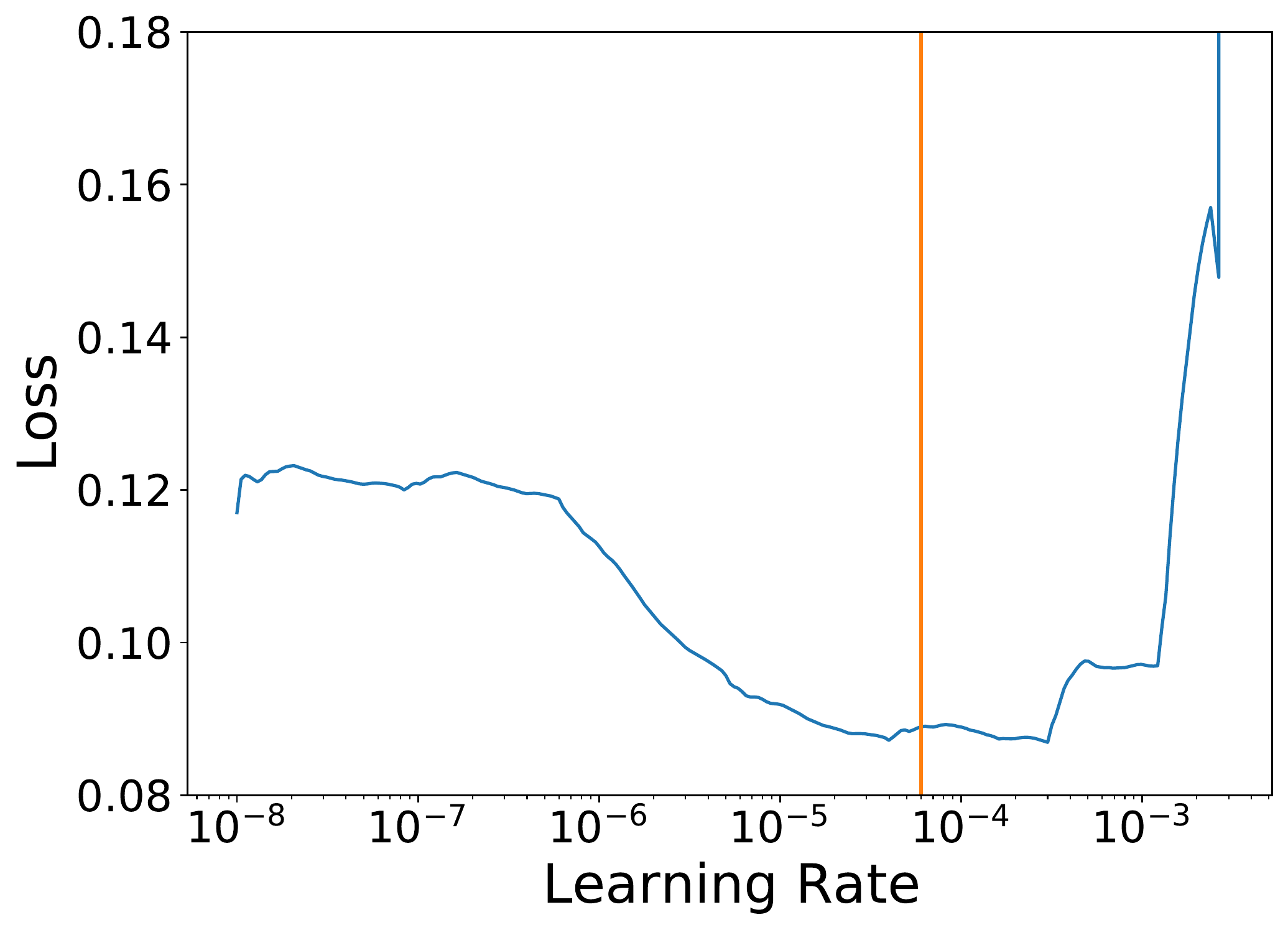}
  \includegraphics[width=49mm]{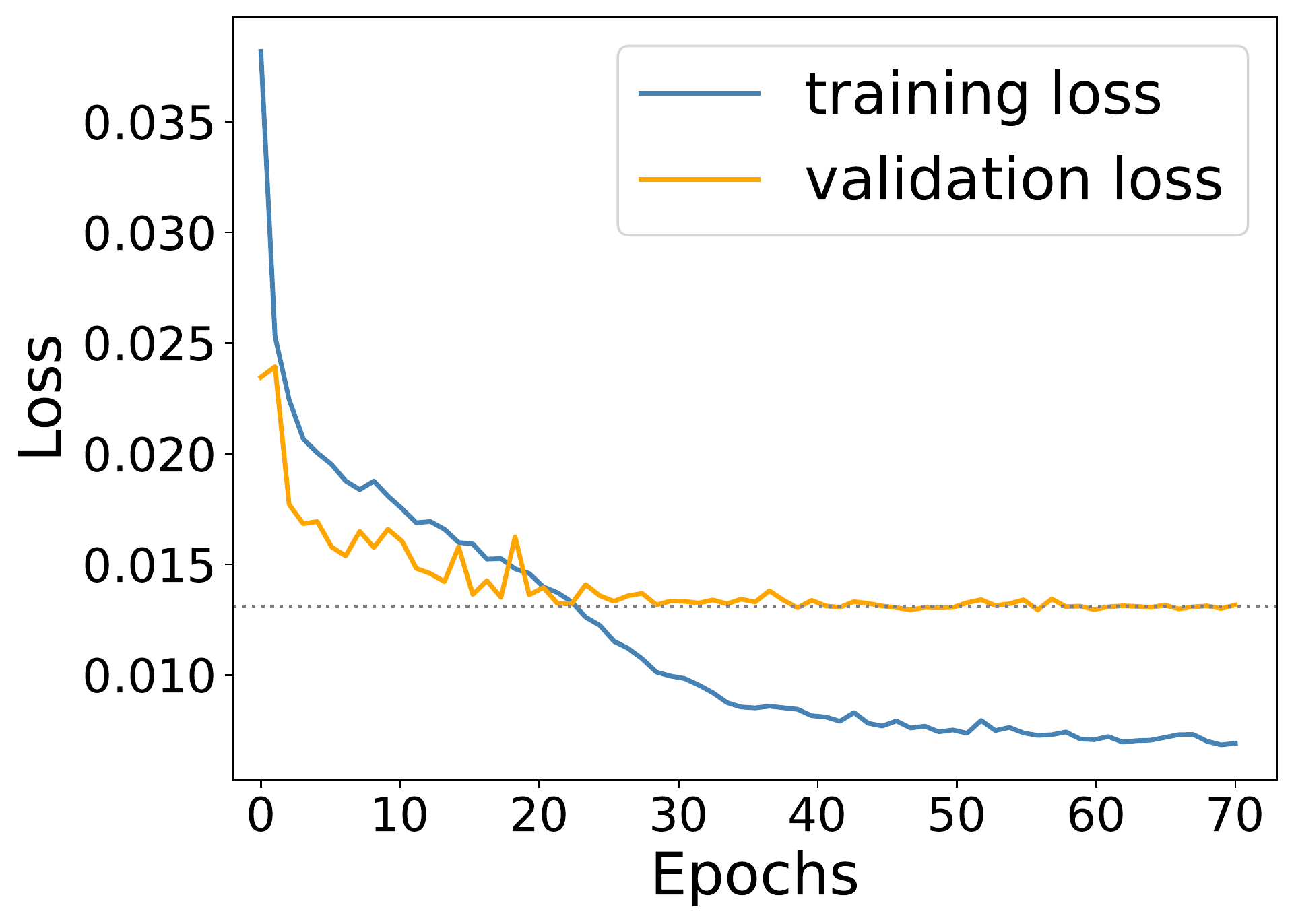}
 \caption{Left: Weights for reconstructing the velocity field $w_u$ as functions of true velocity field. There is no weight applied to the angular part of the velocity loss function.
 Middle: Evolution of loss function as a function of learning rate (blue curve) and the learning rate we have adopted (orange line).
 Right: Loss function of velocity field reconstruction as a function of training epoch.
 }
 \label{figvelweightloss}
\end{figure*}

We find the same outcome in reconstructed the velocity as we did with the density field, as discussed in Section~\ref{sec:recons_dens}. We firstly reconstructed the velocity field by setting $w_{\rm u,i} =1$ in Eq.~\ref{lossvel}. Fig.\ref{fighistw2} shows the comparison between the reconstructed velocity field and true velocity field.  In the bottom panel of Fig.~\ref{fighistw2}, the pink curve shows the difference between the true and reconstructed velocity amplitude values, and there are two 
minimum points in the pink curve at $|{\bf V}_{\rm t}|=115.77$ km s$^{-1}$ ($u_{\rm t}=0.463$) and 
$427.85$ km s$^{-1}$ ($u_{\rm t}=1.7114$).  
Without introducing any specific weight factor $w_u$, the machine learning would try to match only the most common population, which have $V \sim 300~{\rm km~s^{-1}}$. As a result, 
we find that our method 
underestimates the velocities of both the larger and smaller amplitudes. 
Therefore, similar to Section~\ref{sec:recons_dens}, the weight factors are required to reconstruct velocity fields with both large and small amplitudes well.

Similar to what we have done in the previous subsection for the density field, we decompose the true velocity field into three regions  which are bounded by the minimum points of the pink curve of Fig.\ref{fighistw2}. Since the pink curve is  quickly increasing for $u_t>1.7114$, we use a power function model the weights in this region. Finally, we define the weight function as: 
\be \label{fielddd2}
w_{u,i}=\left\{
\begin{aligned}
&c_1, &0< & u_{{\rm t},i} \leq0.463\\
&c_2, &0.463< & u_{{\rm t},i}\leq1.7114\\
&c_3u_{{\rm t},i}^{c_4}+c_5, &1.7114< &u_{{\rm t},i}  \\
\end{aligned}
\right.
\ee
In the above expression, there are five free
parameters $c_1$, $c_2$, $c_3$, $c_4$ and $c_5$. By simultaneously matching the the probability distributions of the reconstructed velocity amplitude values  to the  true
velocity amplitude values in the three regions, we find the optimal choice of the parameters are:   
\be\label{field0}
c_1=0.5,~~c_2=0.18,~~c_3=0.01,~~c_4=5,~~c_5=0.0532.
\ee
The final weight function of Eq.~\ref{fielddd2} is shown in the left-side panel of Fig.~\ref{figvelweightloss}.

The middle panel of Fig.~\ref{figvelweightloss} shows the evolution of the loss function as a function of the learning rate of the Adam
optimizer. 
Similar to Fig.~\ref{figdensweightloss}, if the learning rate is lower than $10^{-6}$, the update of the parameters is too slow, while a too high learning rate ($>2\times10^{-3}$) prevents finding a solution. 
Therefore, we choose the learning rate $6\times10^{-5}$ for velocity field reconstruction. 
The right panel shows the training loss (blue curve) and validation loss (orange curve) by adopting the above learning rate. 
The validation loss tends to be flat beyond epoch 30, and therefore, we stop training at epoch 70. 

\begin{figure*}
\centering
  \includegraphics[width=\columnwidth]{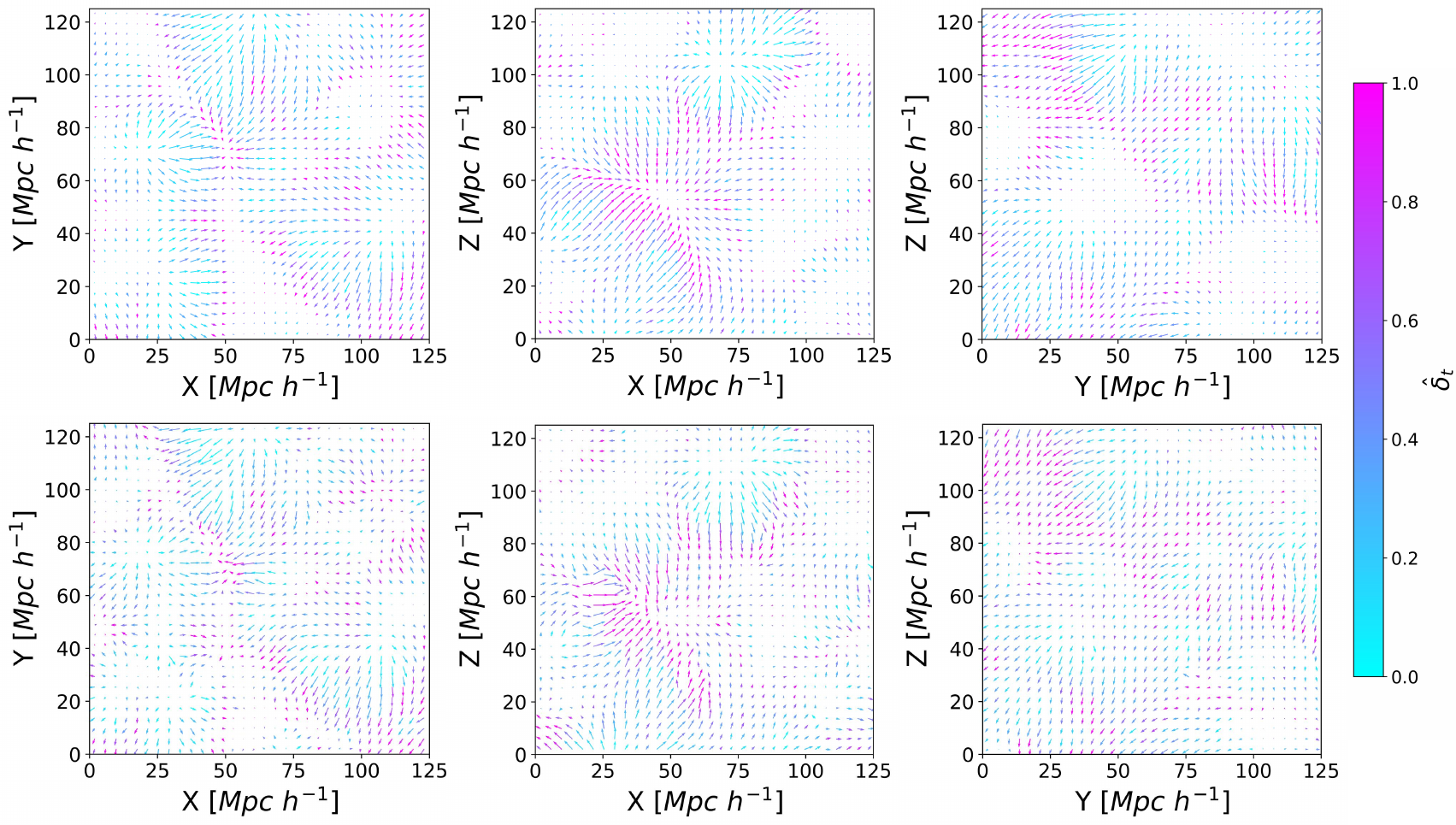}
   \includegraphics[width=\columnwidth]{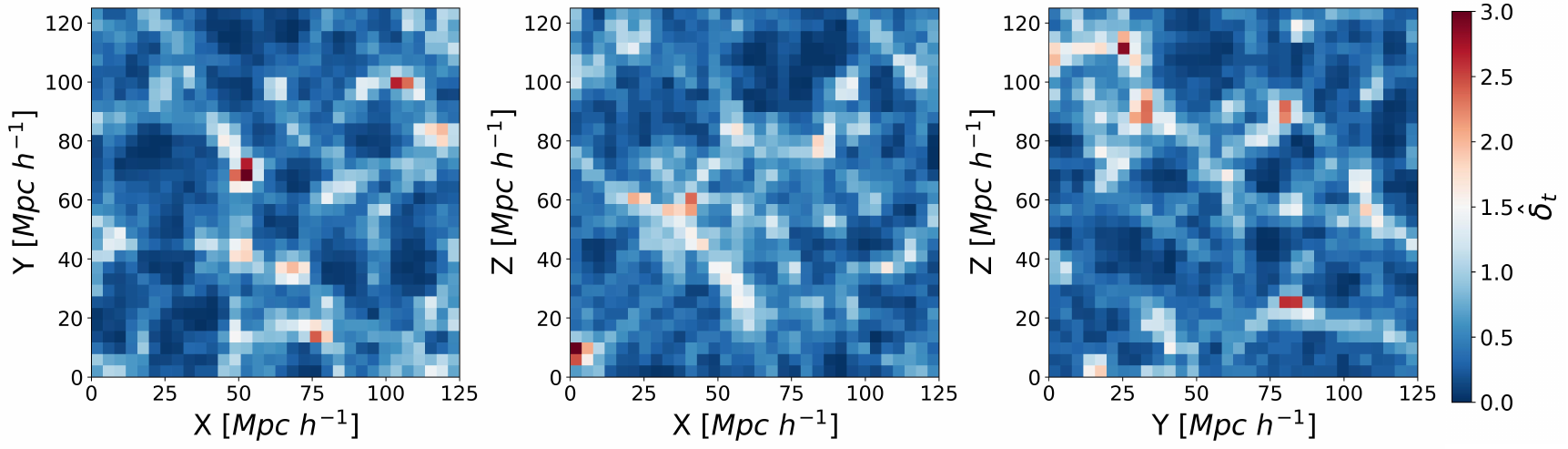}
 \caption{Comparison between the reconstructed velocity fields (top) and the true velocity fields (middle) in XY-, XZ- and YZ-planes, respectively.
 The thickness of each slice is 4 $h^{-1}$~Mpc  (one pixel size).   The bottom panels show the corresponding true density fields.
  } 
 \label{fig8velrecon}
\end{figure*}

Fig.~\ref{fig8velrecon} shows
the comparison between the reconstructed velocity field images (top panels) and  true velocity field images (bottom panels) in the XY-, XZ-, and YZ-planes. Overall, both the amplitudes and directions of the V-net predicted velocity 
excellently match the true velocity, as well as the large-scale structure features. Also, as seen from the velocity fields in the XZ- and YZ-planes (the middle and right panels), there is no significant RSD contamination in the reconstructed real-space velocity fields. 
Note that, however, there is a significant difference on the velocity fields, both in terms of  amplitude 
and direction, at the interface between filaments and massive clusters (XZ-plane with $(X,Z) = (30~h^{-1} {\rm Mpc}, 60~h^{-1}{\rm Mpc})$.
Even with the dark matter-only semi-analytic simulations, such regions might be highly nonlinear and associated with energetic infall shock waves \cite{Hong2014b, Hong2015}.
However, we expect that such local disagreement in nonlinear regions may not significantly affect our goal to estimate the $\beta$ parameter, and we leave it as future works.

\begin{figure*} 
\centering
 \includegraphics[width=54mm]{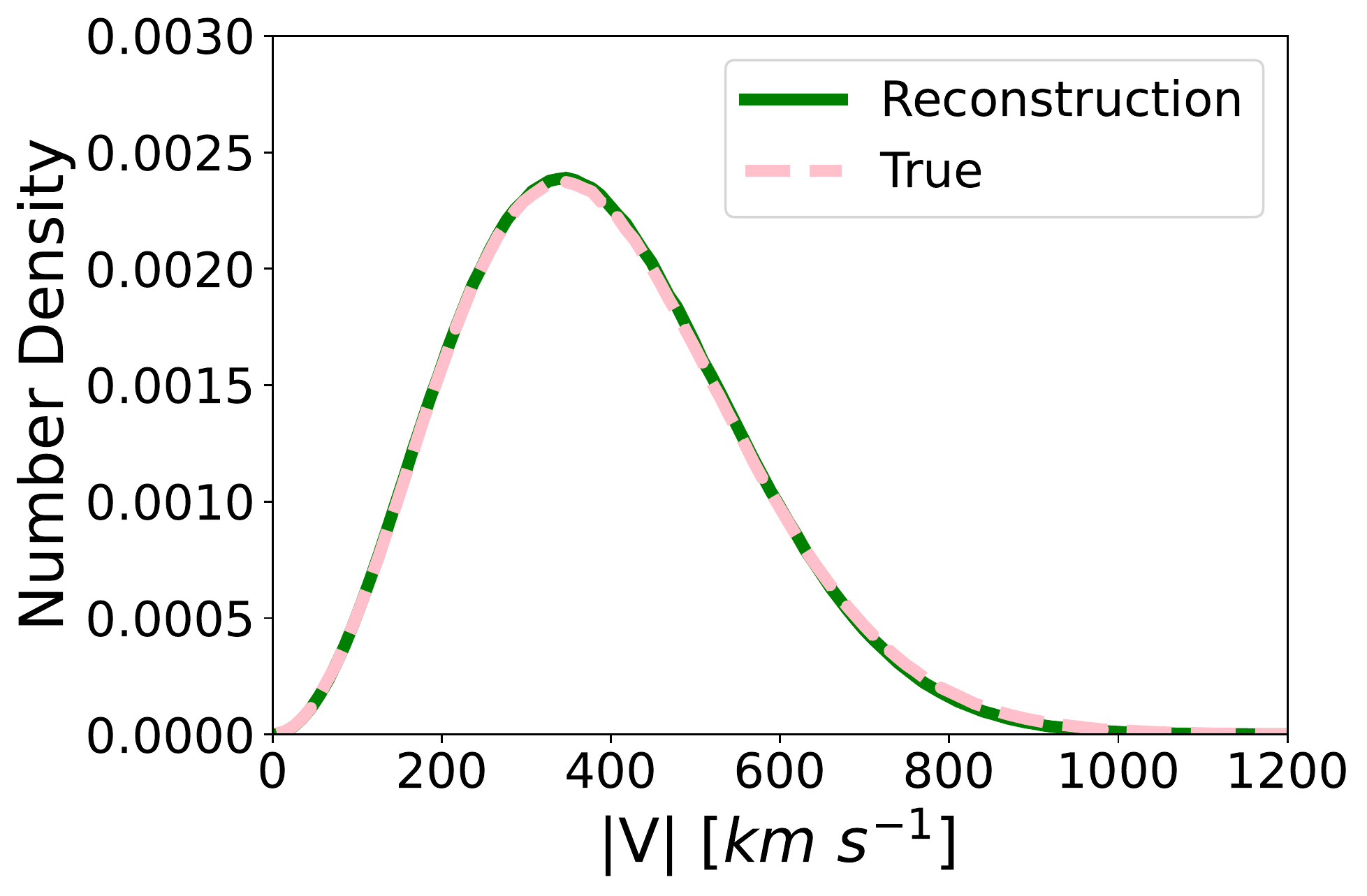}
 \includegraphics[width=48mm]{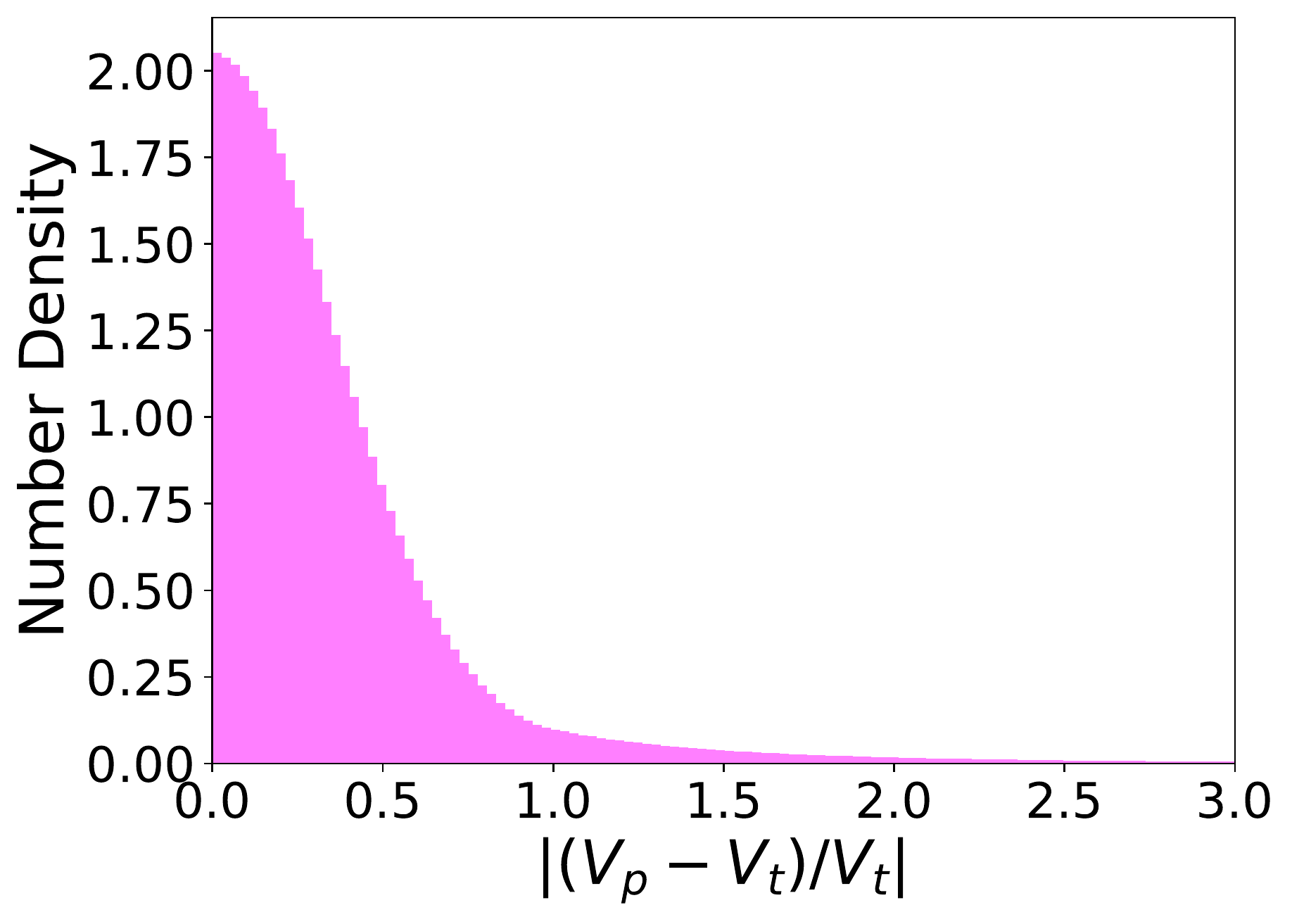}
 \includegraphics[width=48mm]{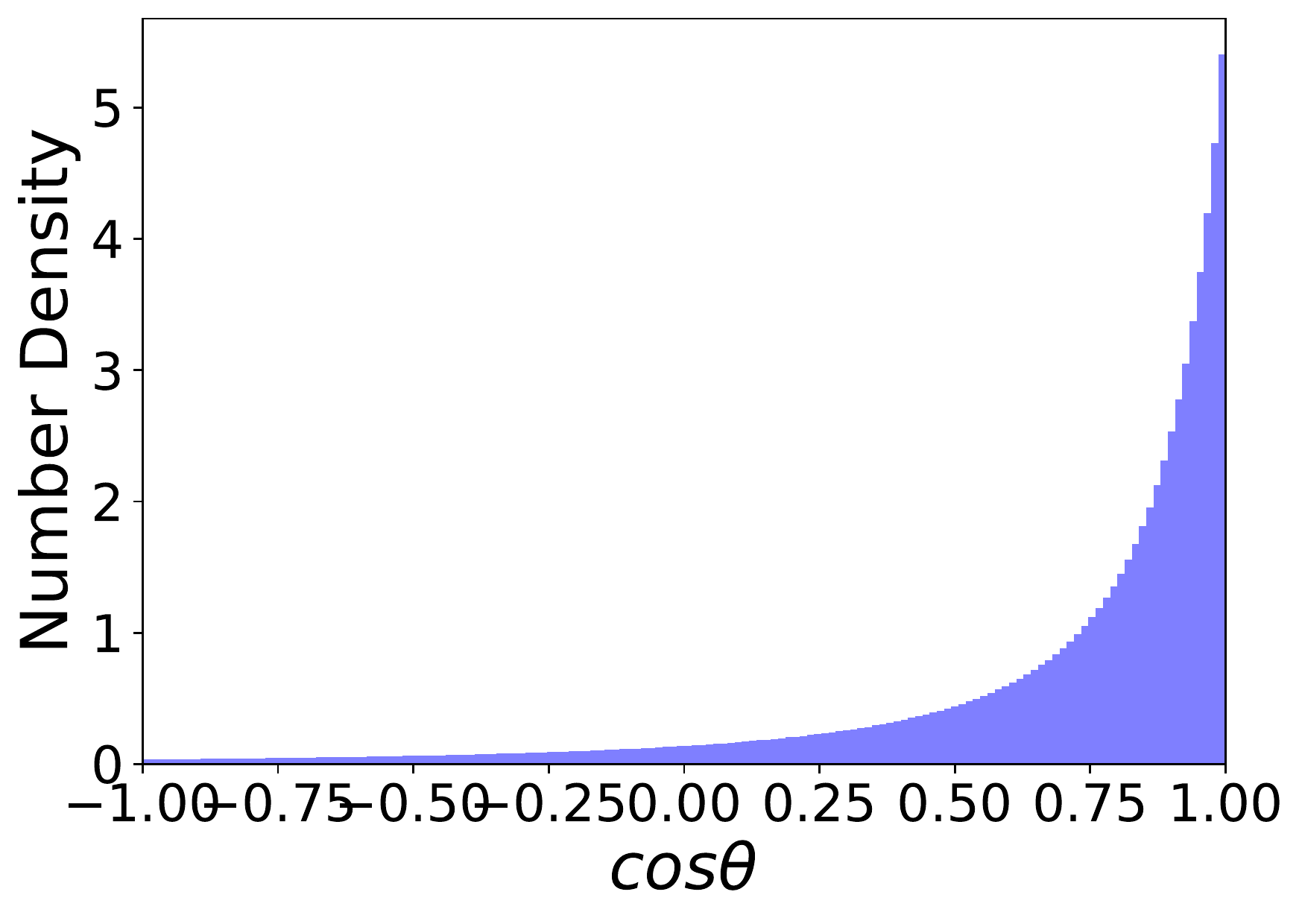}
 \includegraphics[width=\columnwidth]{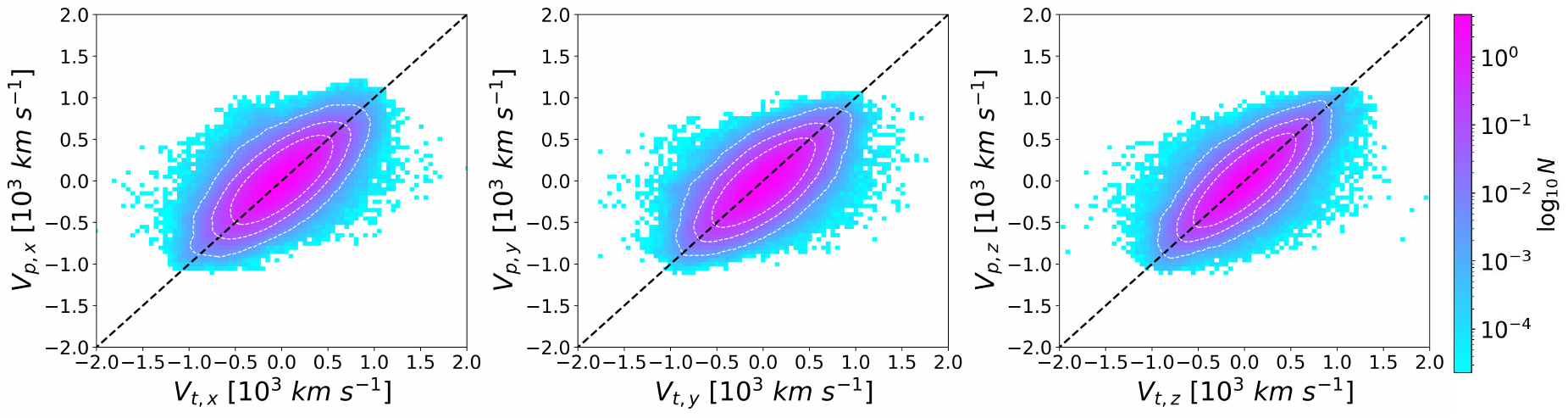}
 \caption{ Top-left: Distributions of the true velocity amplitudes $|{\bf V}_{\rm t}|$ 
 (pink dashed curve) and the reconstructed velocity amplitudes $|{\bf V}_{\rm p}|$ 
 (green solid curve).
 Top-middle: Distribution of the fractional residuals between $|{\bf V}_{\rm t}|$ and $|{\bf V}_{\rm p}|$. 
 Top-right: The distribution of $\cos \theta$. 
 Bottom: Comparisons between the three components of reconstructed velocities to their true velocity counterparts.
 The white colored contours indicate the 68\%, 95\%, and 99.7\%
 confidence levels.}
 \label{fig6histcompv}
\end{figure*}

In the top-left panel of Fig.~\ref{fig6histcompv}, the pink curve shows the distribution of the true velocity amplitudes $V_{\rm t}$, while the green curve shows the distribution of the reconstructed velocity amplitudes $V_{\rm p}$. The top-middle panel displays the fractional residuals between $V_{\rm t}$ and $V_{\rm p}$, most of which are smaller than unity. The top-right panel shows the distribution of $\cos\theta$, where $
\theta$ is the angle between the true and reconstructed velocity vector. In an ideal reconstruction,  most of our reconstructed velocities would have a value close to one $\cos\theta$, and we  find that most of the angles of the velocity vectors do lie close to this limit. 
However, there is a small fraction of the population that have negative values of $\cos \theta$, but we suspect that many of them  have  a velocity amplitude close to zero (see the bottom panel, for example). 

In the bottom panel, we plot the three components of the reconstructed velocities against the true velocities. The white colored contours indicate the 68\%, 95\%, and 99.7\% confidence levels, and the reconstructed velocity fields largely agree with the true velocity fields. 
Interestingly, the distribution of true and reconstructed $V_Z$ is similar to those of $V_X$ and $V_Y$, indicating that our reconstruction does not suffer too much from the effect of contamination from redshift-space distortions.

\begin{figure} 
\centering
\includegraphics[width=85mm]{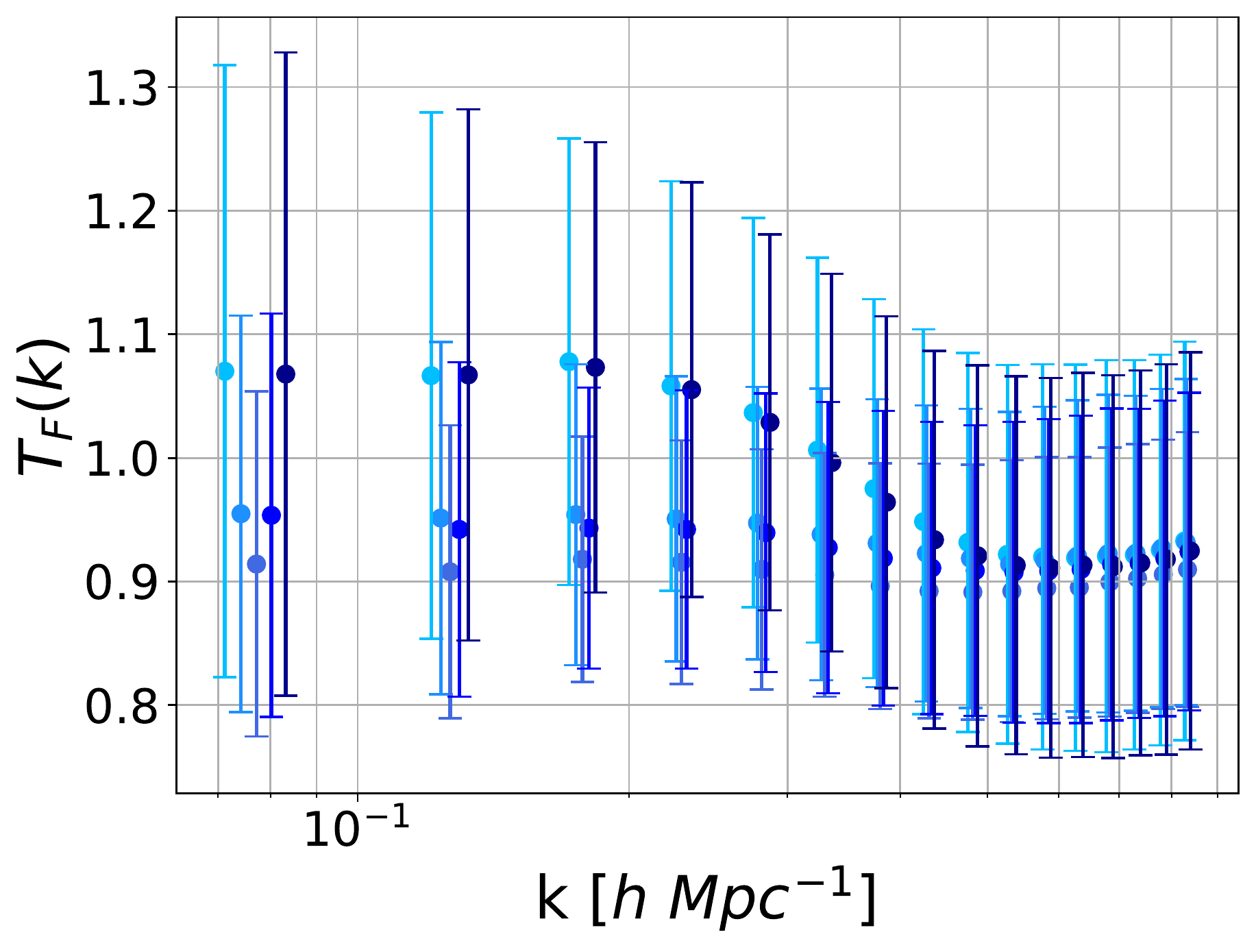}
 \caption{ The transfer function between the power spectrum of the  538 true and 538 reconstructed density fields in five different bins of $\cos\alpha$, where $\alpha$ is the angle between ${\bf V}$ and ${\bf \hat{r}}$. The dots and error bars are the average and the standard deviation of the 538 measurements, respectively. The colors range from light-blue to dark-blue indicating the five bins of $\cos\alpha$, i.e.,  
 $[-1. , -0.6)$, $[-0.6,-0.2)$, $[-0.2,  0.2)$, $[0.2, 0.6)$ and $[0.6,  1 ]$.
}
 \label{figtf}  
\end{figure}

The transfer function, which can be used to quantify the discrepancy between the power spectrum of the divergence of the true and reconstructed velocities, is defined as:
\be 
T_{\rm F}(k)=\sqrt{\frac{P^{\nu\nu}_{\rm pp}(k)}{P^{\nu\nu}_{\rm tt}(k)}}
\ee
where $\nu=\nabla\cdot{\bf V}$ is the divergence of velocity.
In Fig.~\ref{figtf}, we plot the $T_{\rm F}(k)$ in five different bins of $\cos\alpha$, where $\alpha$ is the angular between the line-of-sight (z-direction) and the true velocities. The transfer function values in the five bins all lie close to one, again indicating that our reconstruction does not suffer much from contamination from redshift-space distortions.

\section{Estimating the $\beta$ parameter}\label{sec:beta}

In the utilization of real galaxy surveys, the velocity-velocity comparison method is a   promising method to estimate   $\beta$ using reconstructed velocity fields. This method is the most accurate method used to constrain $\beta$ to date, in comparison to the velocity power spectrum and velocity correlation functions \citep{Carrick2015,Qin2019b,Said2020,Lilow2021}. On the other hand, the correlation functions and power spectrum of the reconstructed velocity fields cannot be used to determine the cosmological parameters, since the reconstructions are the theoretical model rather than velocity measurements\footnote{The velocity measurements should be obtained from the Tully-Fisher relation, fundamental plane and SN-Ia. Then we can measure the power spectrum (or correlation functions) from these measured velocities and compare to the  model power spectrum (or correlation functions) to estimate the cosmological parameters.}.  The goal of this paper is to explore the feasibility of AI reconstructions in terms of the measurements of the cosmological parameters. Here we will test whether or not a reasonable $\beta$ value can be obtained from the reconstructed velocity field using the  velocity-velocity comparison method, to understand if  AI reconstruction is  promising  for real surveys, such as DESI and WALLABY, at a more practical level.  

Once we obtain the reconstructed real-space density and velocity fields, we can compare the reconstructed radial velocities to the measured peculiar velocities to estimate the cosmological parameter $\beta$. Eq.~\ref{betadef}  shows how the velocity field is coupled to the density field through the value of this parameter, which acts as a coupling coefficient.  The reconstructed velocity $V_{\rm p}$ scales with the fiducial $\beta_{\rm fid}=0.518$ of the simulations, while the amplitudes of the measured peculiar velocities of halos $|{\bf V}_{\rm g}|$ should scale with the measured $\beta$. Thus \citep{Springob2014,Springob2016},
\be 
\frac{|{\bf V}_{\rm g}|}{\beta} = \frac{|{\bf V}_{\rm p}|}{\beta_{\rm fid}}~,
\ee 
and therefore, to measure $\beta$, the reconstructed velocity fields should be divided by the fiducial $\beta_{\rm fid}$. 

In observations, the measurement errors of galaxy peculiar velocities are very large  (typically around 20--25\% error on distance estimates for Fundamental Plane and Tully-Fisher relation). To obtain a more accurate estimation of $\beta$, the measurement needs to 
be performed in terms of measured redshift \citep{Springob2014,Springob2016,Carrick2015,Boruah2020}. Therefore, to simulate the measurement procedure in real observations, we 
need to project the reconstructed velocity to the line-of-sight and re-scale to $\beta=1$ using 
\be \label{betswei}
s =\frac{1}{\beta_{\rm fid}}{\bf V}_{\rm p}\cdot\hat{\bf r}~.
\ee 
Then we can calculate the observed redshift $z_{\rm p}$ (of a halo at real-space position ${\bf r}$) predicted from our reconstructions using \citep{Carrick2015,Boruah2020}
\be  
1+z_{\rm p}=(1+z_{\rm h})\left[ 1+\frac{1}{c}(\beta  s + {\bf V}_{\rm ext} \cdot\hat{\bf r}) \right] ~.
\ee 
The reconstruction is usually performed within a limited survey volume $D$, therefore an extra nuisance term  ${\bf V}_{\rm ext}$, namely the residual bulk flow,  is introduced to encapsulate  contributions from beyond the reconstruction volume $D$, and we treat it as a free parameter in our analysis. The Hubble recessional redshift $z_{\rm h}$ is given by \citep{peebles1993,Carrick2015,Boruah2020}:
\be 
z_{\rm h}=\frac{1}{1+q_0}\left( 1-\sqrt{1-\frac{2H_0 r}{c}(1+q_0)} \right) ~,
\ee 
where the deceleration factor is given by $q_0={\Omega_{\rm m}}/{2}-\Omega_{\Lambda}$.

In observations, to accurately estimate the $\beta$ parameter using the velocity-velocity comparison method, both the homogeneous and inhomogeneous Malmquist biases 
are required to be removed. 
The homogeneous Malmquist bias is a consequence
of  selection effects, while 
the inhomogeneous Malmquist bias arises from local density variations due to large-scale structure along the line of sight  \citep{Strauss1995, Springob2016}. A commonly used method to remove these Malmquist biases is the  Forward Likelihood estimation  \citep{Pike2005}.
However, in this paper, we use
simulations with
the exact positions of the  halos, and the Malmquist biases are negligible. Therefore, we use the simple $\chi^2$  minimization to estimate $\beta$. 
The $\chi^2$ between the predicted redshift and the observed redshift of halos is given by \citep{Carrick2015,Boruah2020}:
\be  
\chi^2(\beta, {\bf V}_{\rm ext}) = \sum^{N_{\rm gal}}_{i=1}\frac{(cz_{{\rm obs},i}-cz_{{\rm p},i})^2}{\sigma_v^2} ~,
\ee  
where $z_{\rm obs}$ is given by Eq.~\ref{travp}. The true halo peculiar velocities are used to compute $z_{\rm obs}$, the measurement errors are zeros. Therefore, only the velocity dispersion $\sigma_v$ appears in the above $\chi^2$. The parameters $\beta$ and ${\bf V}_{\rm ext}$ are estimated by minimizing the above $\chi^2$.
 
\begin{figure*}
\centering
\includegraphics[width=75mm]{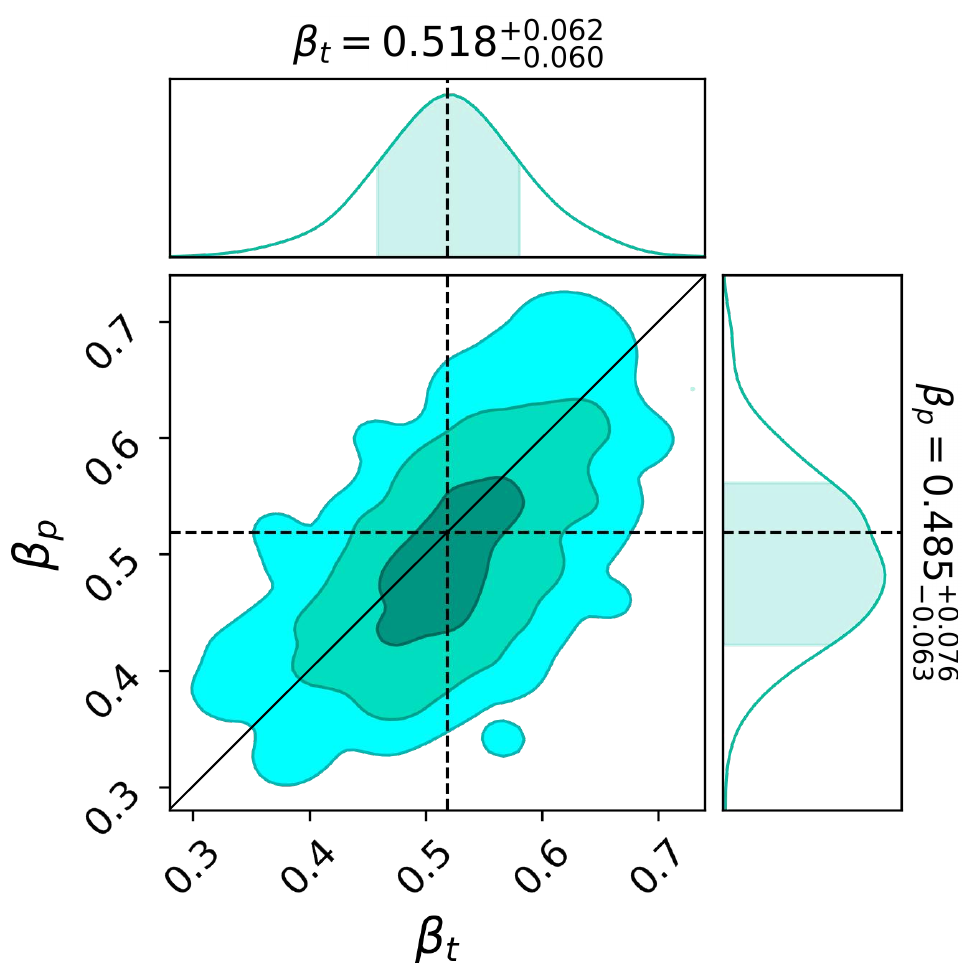}
\includegraphics[width=75mm]{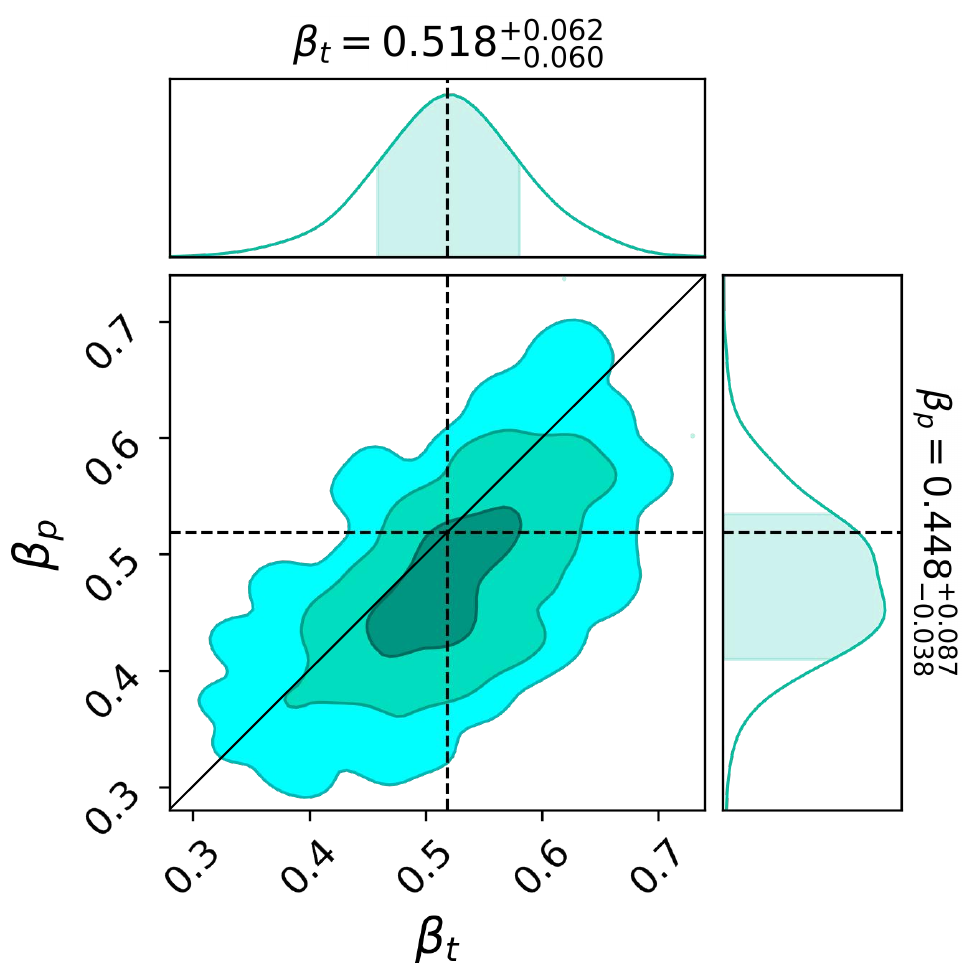}
\caption{Left: Comparison between the $\beta$ measurements from the reconstructed velocity fields ($\beta_{\rm p}$) and those from the true velocity fields ($\beta_{\rm t}$) from the 538 testing sub-cubes.  The weights used for the velocity field reconstruction is set to be $w_{u,i}=1$. 
Top panel shows the probability distribution of $\beta_{\rm t}$. Right side panel shows the probability distribution of $\beta_{\rm p}$, the shade shows the 68\% certainty (1$\sigma$) level. Black dashes indicate the fiducial value $\beta_{\rm fid} = 0.518$. The 2D contours are the 68\%, 95\%, and 99.7\%
certainty levels. Solid line indicates the diagonal ($\beta_{\rm p} = \beta_{\rm t}$).  Right: The same as the Left, but for the weights $w_{u,i}$ of Eq. \ref{fielddd2} and \ref{field0}.  }
 \label{fig9betas}
\end{figure*}

In Fig.~\ref{fig9betas}, we plot the estimations using the reconstructed velocity field, $\beta_{\rm p}$, against the estimations from the true velocity field, $\beta_{\rm t}$, from the 538 testing sub-cubes.
 The left-side panel shows the measurements from the velocity fields reconstructed without weights, i.e., setting weights $w_{u,i}$ to  unity in the loss function of Eq.~\ref{lossvel}. 
The estimated $\beta_{\rm p}$ value ($\beta_{\rm p} = 0.485^{+0.076}_{-0.063}$) agrees with the fiducial value $0.518$ (dashed-lines) within the 68\% confidence level, as shown in the shaded area in the right panel.
While the right-side panel of Fig.~\ref{fig9betas} shows the measurements from the velocity fields reconstructed with weights, i.e., setting the weights $w_{u,i}$ to be given by Eqs.~\ref{fielddd2} and \ref{field0} in the loss function of Eq.~\ref{lossvel}.  The estimated $\beta_{\rm p}$ value ($\beta_{\rm p} = 0.448^{+0.087}_{-0.038}$) agrees with the fiducial value $0.518$ (dashed-lines) within the 68\% confidence level, as shown in the shaded area in the right panel.

However, in both panels of Fig.~\ref{fig9betas}, the $\beta_{\rm p}$ estimated using the reconstructed velocity fields shows a systematic shift compared to the $\beta_{\rm t}$, which is estimated from the true fields ($\beta_{\rm t} = 0.518^{+0.062}_{-0.060}$).  The scatter in the estimated values of $\beta_{\rm p}$ are result of the variance between the sub-cubes and the whole simulation box, as well as the variance between the V-net reconstructed fields and the true fields.  
 
\begin{figure*}
\centering
\includegraphics[width=75mm]{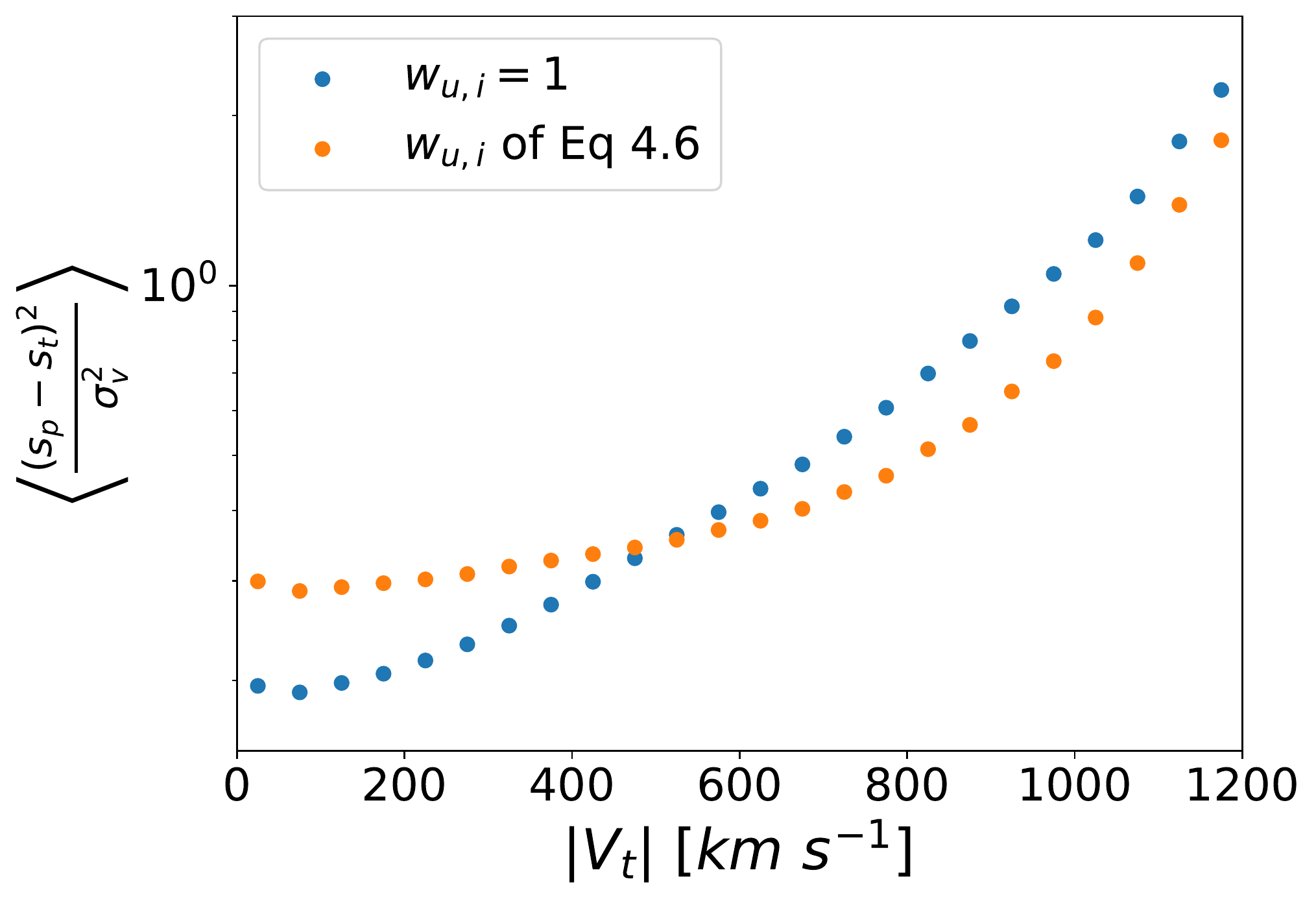}
\caption{ The normalised velocity residual for the peculiar velocity halo catalogues, as a function of the true velocity amplitude. The  reconstruction with (\textit{orange}) weights emphasises better training for higher velocity magnitudes, leading to a smaller residual compared to the reconstruction without (\textit{blue}) weights at high velocity amplitudes. However, the use of weights leads to a bias at low velocities, leading to a higher residual, compared to with weights set to unity.}
 \label{fig9beta2s}
\end{figure*}

We also find that, when using the velocity-velocity comparison method to estimate $\beta$, the velocity reconstructions without weights give more accurate $\beta$ estimations, comparing to the velocity reconstructions with weights, although the velocity reconstructions with weights yield  a more precise estimate and tighter probability distributions. This is mainly due to the fact that the $\chi^2$ used in the velocity-velocity comparison method reflects the difference between the reconstructed velocities and the `true' velocities, i.e., the velocities  that are directly inferred from  the true velocities of the dark matter particles. So, minimizing the $\chi^2$ is (to some extent) equivalent to minimizing the loss function without weights, which  automatically tries to minimize the all velocity amplitude   and angular differences between true and predicted velocity fields. However, the introduction of weights may bias this $\chi^2$ calculation, as shown in \ref{fig9beta2s}, since data points from different velocity regions will be treated differently (a higher weight meaning greater priority to fitting the highest and low velocity amplitudes during training). Therefore, the reconstructions without weights give a smaller $\chi^2$ and so more accurately estimate the $\beta$ values.

\begin{figure*}
\centering
\includegraphics[width=50mm]{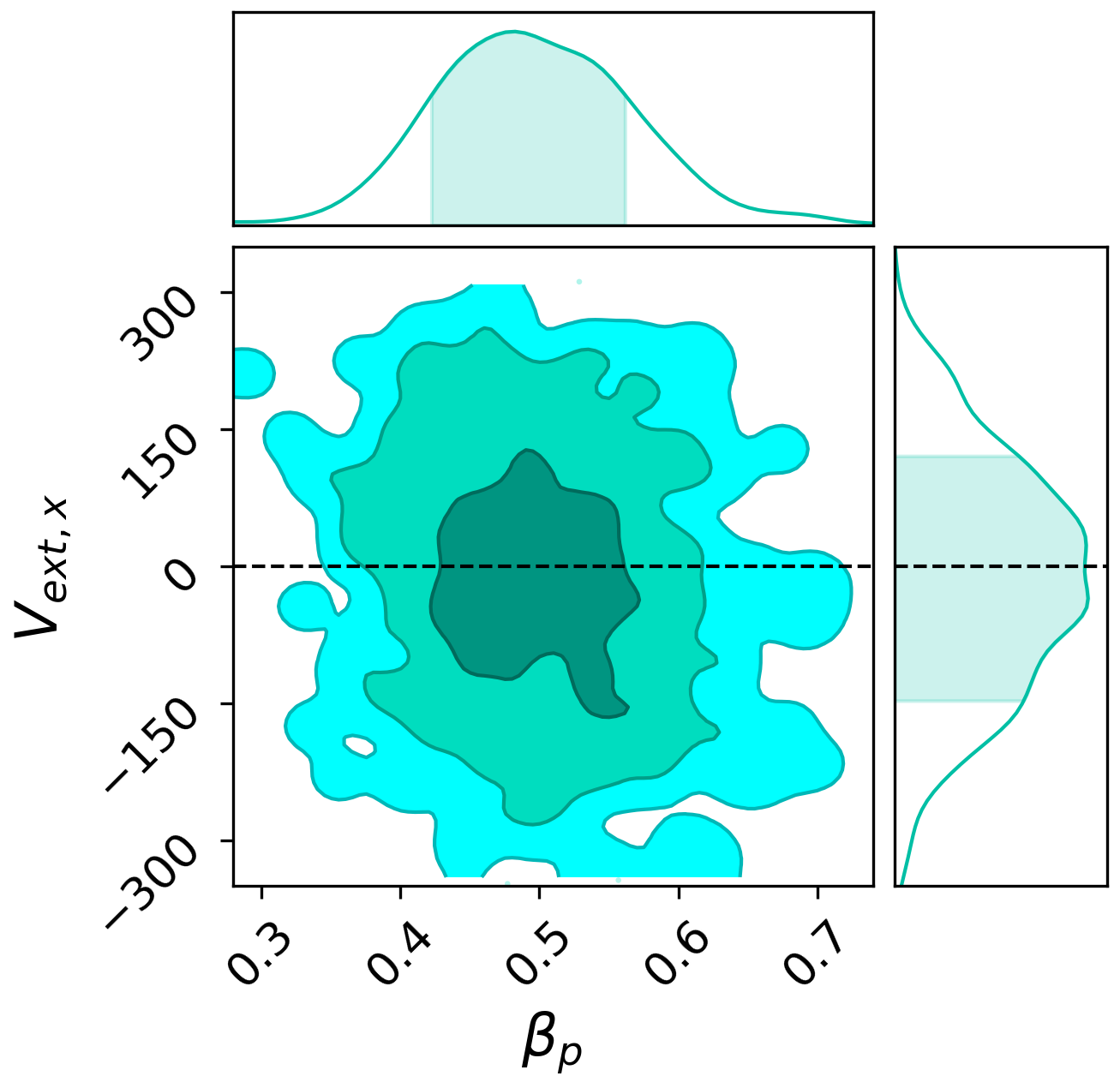}
\includegraphics[width=50mm]{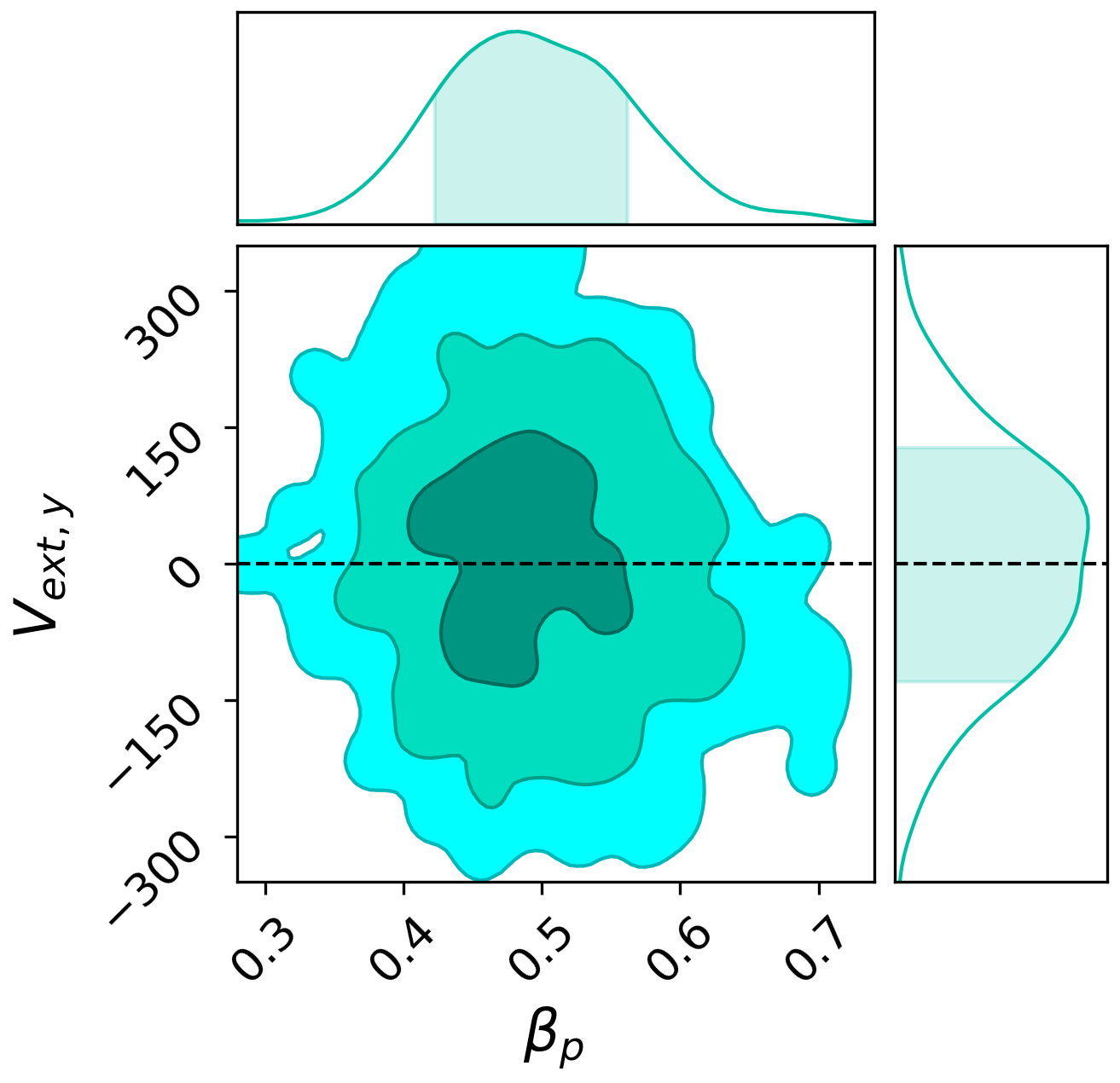}
\includegraphics[width=50mm]{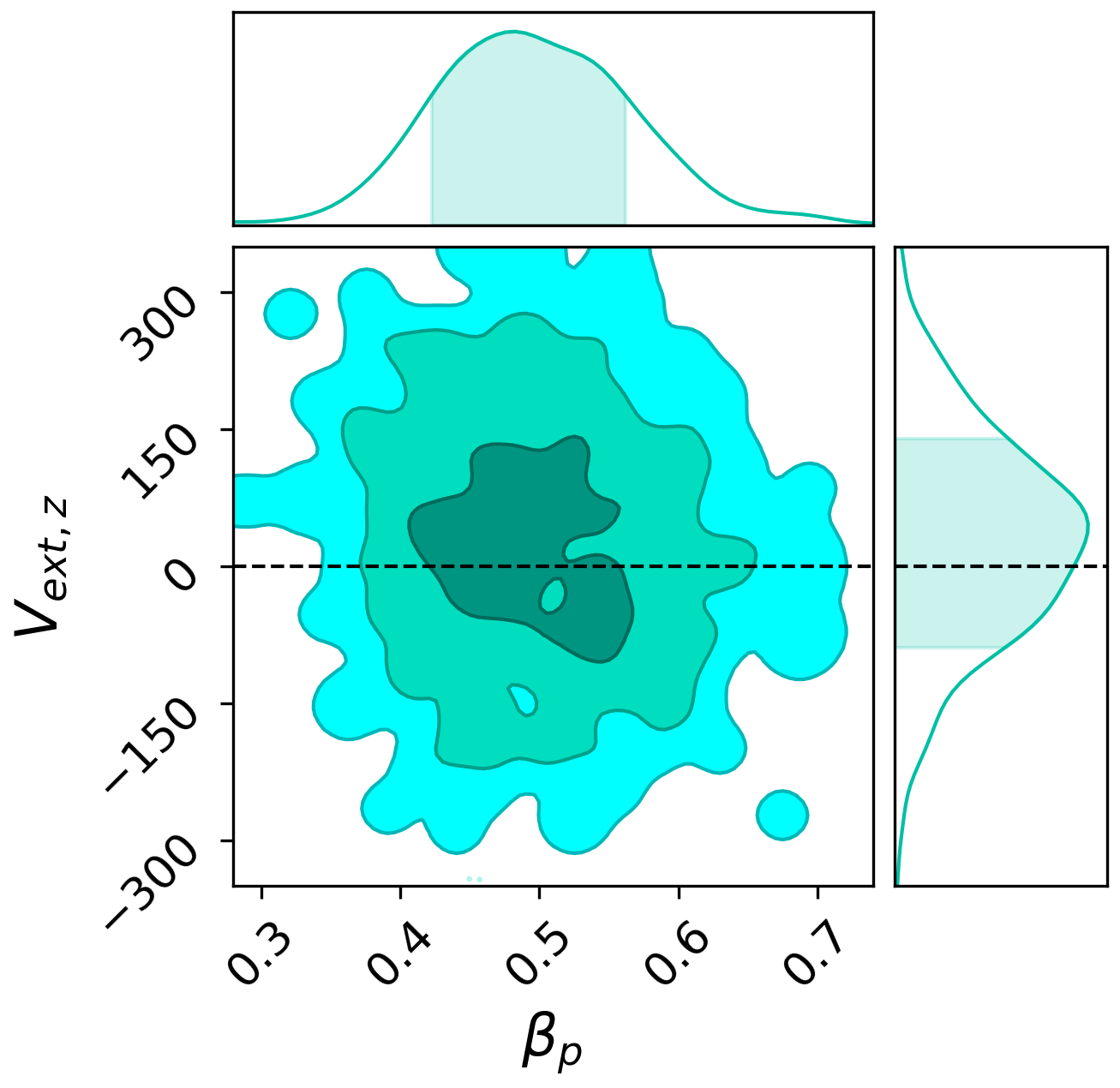}
\includegraphics[width=50mm]{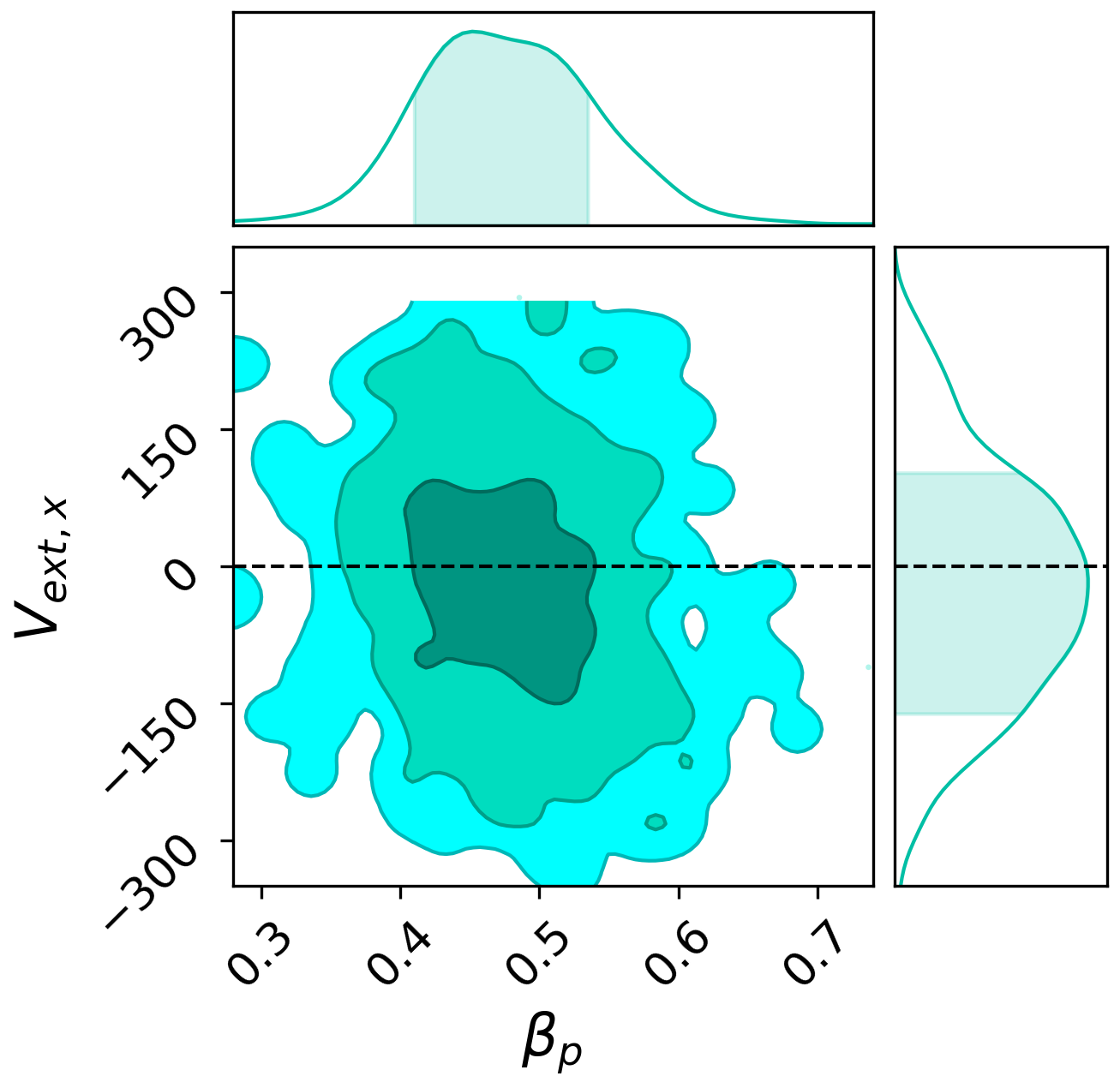}
\includegraphics[width=50mm]{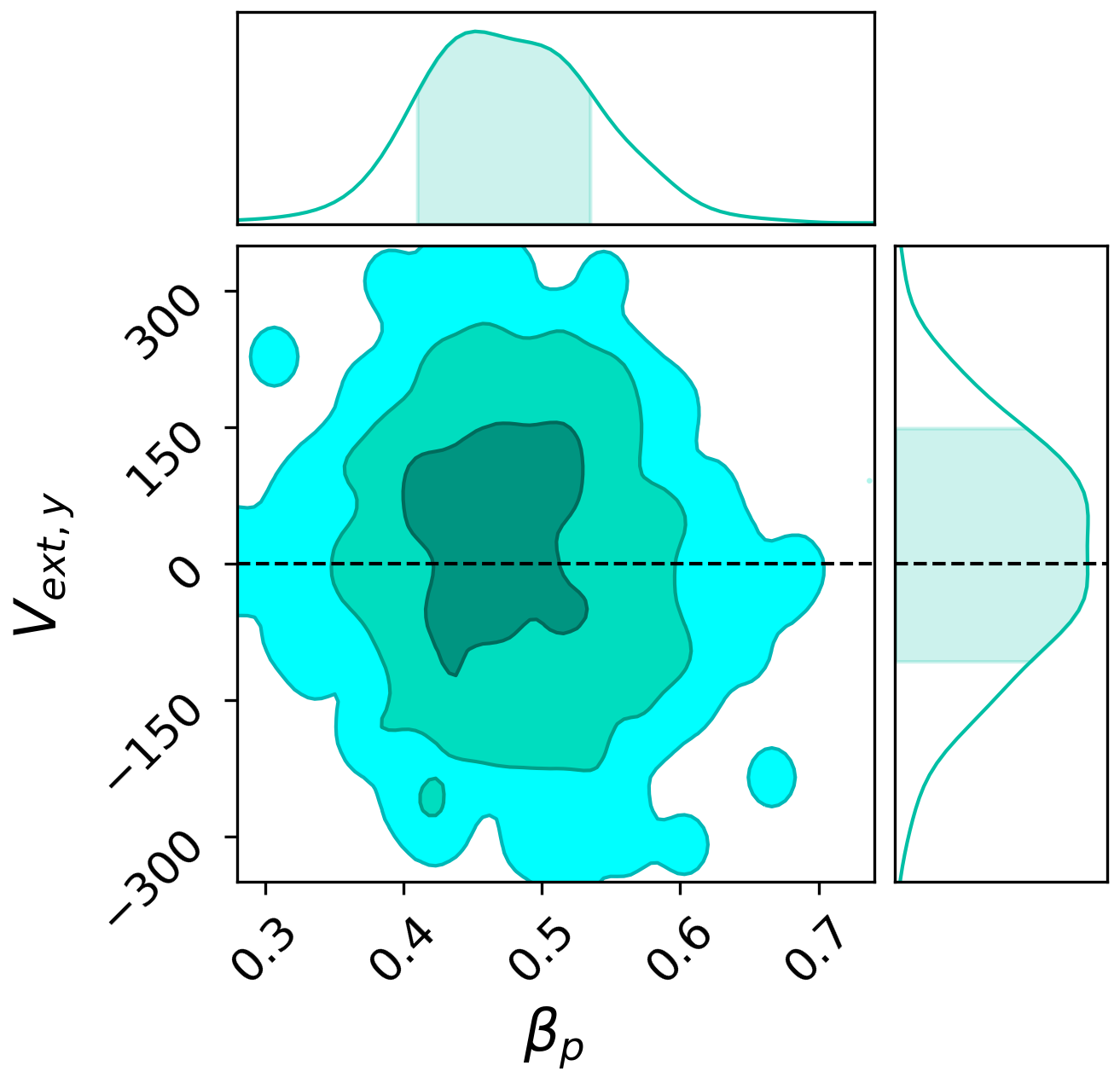}
\includegraphics[width=50mm]{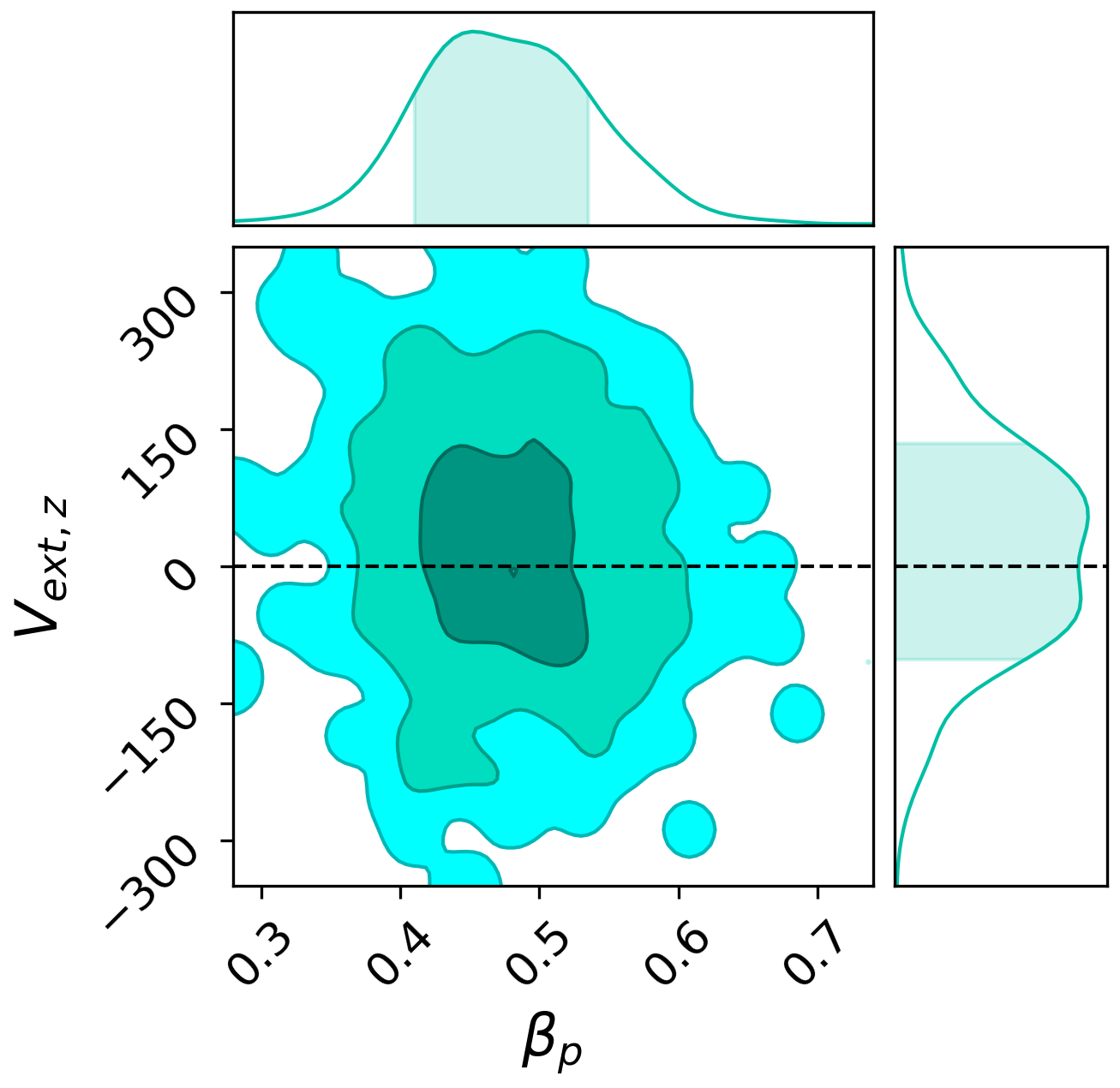}
\caption{  Top: The ${\bf V}_{\rm ext}$ measurements from the reconstructed velocity fields with weights $w_{u,i}=1$. Bottom: The ${\bf V}_{\rm ext}$ measurements from the reconstructed velocity fields with weights set to be Eqs.~\ref{fielddd2} and \ref{field0}. }
 \label{fig14vs}
\end{figure*}

In Fig.~\ref{fig14vs}, we plot the estimations of ${\bf V}_{\rm ext}$ against $\beta_{\rm p}$, from the 538 testing sub-cubes.
There are no correlations between $\beta$ and ${\bf V}_{\rm ext}$. The estimated values of ${\bf V}_{\rm ext}$ are, as expected, centred at zero.

\section{Conclusion}\label{sec:conc}

The peculiar velocities of galaxies serve as a valuable tool for studying the gravitational potential field, allowing for a more direct measurement of the large-scale distribution of matter than positions of galaxies alone. However, effectively utilizing these velocities requires a precise prediction of the density and velocity fields. In this study, 
we use the simulated halos
as proxies for the galaxies, and
we evaluate the effectiveness of a Convolutional Neural Network (CNN) architecture, the V-net, in reconstructing these fields and making inferences about the growth rate of cosmological structure.

In this work, the real-space density fields are reconstructed from input halo redshift catalogs using an initial V-net. Results indicate that tuning the loss function with density-dependent weights enhances the V-net's accuracy in predicting  both the probability distribution of density values (i.e., density values across both low- and high-density regions) and the cosmological features (i.e., two point correlation function) of the density field. Additionally, the reconstructed density fields exhibit minimal contamination from redshift-space distortions in the input halo catalogs.

The velocity fields are then reconstructed from the density fields using another V-net. The results demonstrate that tuning the loss function with velocity-dependent weights enhances the accuracy of the V-net in predicting pixels with large velocity amplitudes and the probability distribution of velocity amplitude values. As with the reconstructed density fields, the velocity fields also exhibit reduced contamination from redshift-space distortions in the input halo catalogs.

The estimation of the cosmological parameter $\beta$ through comparison of the reconstructed and true velocity fields yields results consistent with the fiducial value. While the $\beta$ values estimated using the reconstructed velocity fields exhibit a systematic offset relative to the true fields, they still agree with the fiducial value within $1\sigma$. This suggests that the density and velocity fields reconstructed using AI provide reasonably accurate constraints. The $\beta$ values estimated using the reconstructed velocities without weights are more accurate than the reconstructions with weights, as there is better matching (in terms of angle and amplitude) for the velocities of the peculiar velocity halos across the range of possible velocities. This better matching leads to a smaller $\chi^2$ value for the velocity-velocity comparison method, and a more accurate fit to the data.

This study successfully illustrates the concept of reconstructing large-scale density and velocity fields through the use of a CNN architecture. However, its scope is limited, as it only considers a cubic geometry without observational selection effects, simplistic peculiar velocity errors, and only a single cosmological model. To address these complexities, additional simulations and training would be necessary, and such efforts are left for future studies.  We also plan to use mock galaxy catalogues (generated using a halo-occupation distribution or semi-analytic model)   rather than halo catalogues for the training and testing of the CNN. Our ultimate goal will be the application of the trained CNN to real surveys (e.g. WALLABY and DESI), to reconstruct the density and velocity fields in the nearby Universe, and also improve estimates of the $\beta$ parameter at higher redshifts.


\acknowledgments

FQ, DP, and SEH are supported by the project \begin{CJK}{UTF8}{mj}우주거대구조를 이용한 암흑우주 연구\end{CJK} (``Understanding Dark Universe Using Large Scale Structure of the Universe''), funded by the Ministry of Science. CGS acknowledges support via the Basic Science Research Program from the National Research Foundation of South Korea (NRF) funded by the Ministry of Education (2018R1A6A1A06024977 and 2020R1I1A1A01073494). This work was supported by the high-performance computing cluster Seondeok at the Korea Astronomy and Space Science Institute.

We would like to acknowledge the Korean Astronomy and Machine Learning (KAML) meeting series for providing a forum for fruitful discussions on the intersection of astronomy and machine learning.

\bibliography{FQinRef}{}
\bibliographystyle{aasjournal}






\end{document}